\documentclass{article}

\usepackage{amsmath,amsfonts,amsthm,cite}

\numberwithin{equation}{section}

\renewcommand{\Re}{\mathop{\mathrm{Re}}}

\newtheorem{thm}{Theorem}
\newtheorem{lem}[thm]{Lemma}
\newtheorem{prop}[thm]{Proposition}
\newtheorem{corol}[thm]{Corollary}

\theoremstyle{remark}

\newtheorem*{rem}{Remark}

\newcommand{\spec}[1]{\mathop{\mathrm{spec}}(#1)}
\newcommand{\res}[1]{\mathop{\mathrm{res}}(#1)}

\DeclareMathOperator{\supess}{sup\,\,ess}
\DeclareMathOperator{\supp}{supp}
\DeclareMathOperator{\sign}{sign}

\newcommand{\RR}{\mathbb{R}}
\newcommand{\CC}{\mathbb{C}}

\begin{document}

\pagestyle{myheadings}
\thispagestyle{empty}

\title{\bf Continuity properties of integral kernels\\
associated with
Schr\"odinger operators on manifolds}

\author{Jochen Br\"uning \dag, Vladimir Geyler \ddag, Konstantin Pankrashkin \dag\S\\[\bigskipamount]
\dag{} Institut f\"ur Mathematik, Humboldt-Universit\"at zu Berlin,\\
Rudower Chaussee 25, 12489 Berlin, Germany\\[\smallskipamount]
\ddag{} Mathematical Faculty, Mordovian State University,\\ 430000 Saransk, Russia\\[\smallskipamount]
\S{} Corresponding author, E-mail: const@mathematik.hu-berlin.de}

\date{}

\maketitle

\begin{abstract}
For Schr\"odinger operators (including those with magnetic fields)
with singular scalar potentials on manifolds
of bounded geometry, we study continuity properties of some
related integral kernels: the heat kernel, the Green function, and
also kernels of some other functions of the operator. In
particular, we show the joint continuity of the heat kernel and
the continuity of the Green function outside the diagonal.  The
proof makes intensive use of~the~Lippmann--Schwinger equation.
\end{abstract}


\setcounter{section}{-1}

\section{Introduction}

The analysis of Schr\"odinger operators occupies a central place
in quantum mechanics. Suitably normalized, over the configuration
space $\RR^n$ these operators have the form
\begin{equation}
            \label{HAU1}
H_{A,U}=(-i\nabla-A)^2+U,
\end{equation}
where $A$ is the magnetic vector potential and $U$ is an electric
potential. A~huge literature is dedicated to the study of
properties of $H_{A,U}$ in its dependence on $A$ and $U$, see
the recent reviews \cite{Sim20,Sim21,RM}. An essential feature of the
quantum-mechanical operators in comparision to the differential
operator theory is admitting singular potentials~\cite{FLS},
although the operator itself preserves some properties like
regularity of solutions~\cite{Hinz}.

In generalizing the Euclidean case it is natural to consider
operators acting on curved spaces like Riemannian manifolds, where
the operators take the form
\begin{equation}
H_{A,U}=-\Delta_A+U \label{HAU2}
\end{equation}
with $-\Delta_A$ being the Bochner Laplacian. It is worthnoting
that the study of the quantum-mechanical Hamiltonians on
Riemannian manifolds goes back to Schr\"odinger \cite{Sch} and is
not only of mathematical interest. Besides the applications to
quantum gravity and to other fields of quantum physics where
geometrical methods play a crucial role, properties of the
Schr\"odinger operators on curvilinear manifolds find extensive
applications in contemporary nanophysics, see, e.g.~\cite{CMR,GK}.

One of the important questions in the investigation of the
Schr\"odinger operators is the continuity properties of related
integral kernels, for example, of the Green function $G(x,y;z)$ or
of the heat kernel. Many physically important quantities are
expressed through the values of these kernels at some points or
their restrictions onto submanifolds, and these values are
meaningless or, strictly speaking, are not defined if the kernels
are not continuous (as integral kernels are, generally speaking,
only measurable). For example, the calculation of the so-called
Wigner $\mathcal R$-matrix involves the values $G(a,b;z)$ of the
Green function at certain fixed $a$ and $b$; matrices of such form
are widely used in the theory of zero-range
potentials~\cite{AGHH}, the scattering theory~\cite{WE} and the
charge transport theory \cite{WKRS}. Other examples are provided
by the heat kernel trace used in the quantum
gravity~\cite{Avr,Vas} or by the calculation of the density of
states involving the trace of the renormalized Green function,
i.e. its suitably renormalized restriction onto the
diagonal~\cite{LGP}. We also note that the continuity of the Green
function frequently implies  {\it a priori} the continuity of the
eigenfunctions. We remark that the differentiability of the
kernels is also of interest in some problems~\cite{KSB}.

For the Schr\"odinger operator \eqref{HAU1} without magnetic
vector potential ($A=0$) acting over a Euclidean configuration
space, the continuity of the naturally related integral kernels
was proved by B.~Simon~\cite{Sim} for (singular) potentials from
the Kato classes. The continuity in the case of the presence of
magnetic vector potentials was stated in~\cite{Sim20} as an open
problem; only several years ago Simon's results were extended
in~\cite{BHL,BHM} to magnetic Schr\"odinger operators on domains
in Euclidean space with vector and scalar potentials of Kato's
type. In the both cases, the proof used certain probablistic
technique. A part of the results concerning bounds of for the
kernels admits a generalization to elliptic operators with
singular coefficients of a more general form, see e.g.
\cite{DH,GerK,VK,LSem,LS,Ouh,Yaj}.

The theory of Schr\"odinger operators with singular potentials on
manifolds is still far from complete. There are numerous works
concerning the bounds for the heat
kernels~\cite{BCG,TC1,Gri1,Grig2,Grig3,GSC,ZZ}, mapping properties
\cite{YK1,YK2,Shu,Skr} or some particular questions of the
spectral analysis~\cite{Ouh2,Pou,Shen,KSh}, but the attention has been
mostly concentrated either on the free Laplace-Beltrami operator
or on special potentials. Sufficiently wide conditions for the
essential self-adjointness of the Schr\"odinger operators have
been established only very recently~\cite{BMS,Shu1} (see
also~\cite{Mil1,Mil2} for further developments). In this paper we
are interested in the regularity properties of the kernels rather
than the bounds for them; these questions were not addressed in any
of the previous works. Because of the above described situation in
the study of Schr\"odinger operators on manifolds, our
restrictions on vector and scalar potentials are slightly stronger
than in the Euclidian case, and they are of  a different nature:
the Kato or Stummel classes used in~\cite{Sim} have some
relationship to the probabilistic technique, while our conditions
come mostly from the operator-theoretical methods and formulated
in terms of $L^p$-spaces (see Subsection~\ref{cond-pot} below).
Nevertheless, the class of potentials  we consider is wide enough
in order to include physically reasonable local singularities and
to satisfy the requirement stated by B.~Simon~\cite{Sim}, as it
includes  all continuous functions and Coulomb-like local
singularities.

As it was mentioned already, the study of Schr\"odinger operators
in the Euclidean case involved some probabilistic tools like
Brownian motion or the Feynman--Kac formula. Here we employ a
completely different technique from operator theory. Our main tool
is the Lippmann--Schwinger equation for self-adjoint operators $A$
and $B$ with common domain:
\[
(A-\lambda)^{-1}-(B-\lambda)^{-1}=(B-\lambda)^{-1}(A-B)(A-\lambda)^{-1}\,.
\]
If $(A-\lambda)^{-1}$ and $(B-\lambda)^{-1}$ are integral
operators, so is the right-hand side, but its kernel tends to have
better regularity properties then both of the kernels on the left.
Such an observation being combined with arguments like elliptic
regularity provides the continuity of the Green function, which
can be transferred to other kernels (in particular, to the heat
kernel), using a combination of operator methods from
\cite{CFKS,Sim}.

We would like to emphasize that, in contrast to the probabilistic
technique, our approach can be applied to higher order
differential operators. Moreover, the higher the order of an
elliptic operator, the easier it is to satisfy the conditions of
the main lemma~\ref{lemma3}, so that our methods can give new
results also in the Euclidean case. Nevertheless, we restrict
ourselves to Schr\"odinger operators on manifolds in this paper.

The paper is organized as follows. In Section~\ref{sec1}, we
collect some facts about Schr\"odinger operators on manifolds of
bounded geometry and introduce the class potentials $A$ and $U$
in~\eqref{HAU2} to deal with. Section~\ref{sec2} contains some
important integral estimates. In Section~\ref{sec3}, we derive
some estimates for the resolvent norms as well as necessary bounds
for the heat kernel. Section~\ref{sec4} is devoted to the proof of
the main result, Theorem~\ref{main}, which contains the continuity
of integral kernels for various functions of the operator. In the
last section, Section~\ref{sec5}, we discuss briefly possible
generalizations and perspectives.

\section{Preliminaries}\label{sec1}

\subsection{Geometry}
By $X$ we denote a complete connected Riemannian manifold with
metric $g=(g_{ij})$. Throughout the paper we suppose that $X$ is
of bounded geometry, which means that the injectivity radius
$r_\mathrm{inj}$ of $X$ is strictly positive and all the covariant
derivatives of arbitrary order of the Riemann curvature tensor are
bounded. Examples are provided by homogeneous spaces with
invariant metrics, compact Riemannian manifold and their covering
manifolds, the leaves of a foliation of a compact Riemannian
manifold with the induced metric; we refer to~\cite{Roe},
\cite{Shu} for further examples and a more extensive discussion.
We put $\nu:=\dim X$; through the paper $d(x,y)$ denotes the
geodesic distance between points $x,y\in X$, the open ball with
center $a\in X$ and radius $r$ is denoted by $B(a,r)$,
$D=\{(x,y)\in X\times X\,:\,x=y\}$ denotes the diagonal in
$X\times X$. The integral of a function $f$ on $X$ with respect to
the Riemann--Lebesgue measure on $X$ is denoted by
$\displaystyle\int_X\,f(x)\,dx$, and $V(a,r)$ denotes the
Riemannian volume of $B(a,r)$. We also fix a number $r_0=r_0(X)$,
such that $0<r_0<r_{\rm inj}$. The following properties of
manifolds with bounded geometry will be used below (see e.g.
Sections A1.1 and 2.1 in~\cite{Shu}, as well as \cite{Tri} for
proofs and additional bibliographical hints).
\begin{description}
\item[(V1)] There is a constant $w_1\ge1$ such that for every
$a,b\in X$ and $0<r\le r_0$
\[
w_1^{-1}\le \frac{V(a,r)}{V(b,r)}\le w_1\,.
\]
\item[(V2)] There are constants $w_2>0$ and $\theta_X>0$ such that
for all $a\in X$ and $r>0$
\[
V(a,r)\le w_2\,e^{\theta_X r}.
\]
\item[(V3)] There is a constants $w_3\ge1$ such that in each ball
$B(a,r_0)$ there holds $w_3^{-1}\le\sqrt{\det[g_{ij}(x)]}\le w_3$
with respect to the normal coordinates $x$ in $B(a,r_0)$.

\end{description}
Put $V_s(r):=\sup_{x\in X}
\,V(x,r)$, $V_i(r):=\inf_{x\in X}\,V(x,r)$. Then the
properties (V1) and (V2) imply
\begin{description}
\item[(V4)] $0<V_i(r)\le V_s(r)<\infty$\,\,\,\,$\forall\,r>0$,
\item[(V5)] $V_s(r)=O(r^\nu)$ as $r\to 0$.
\end{description}

\noindent Moreover, from the well-known Toponogov triangle
comparison theorem (see, e.g. \cite{Ber}, p.~281) we have

\begin{description}
\item[(V6)] If $f_a$ denotes the inverse of the exponential map in
$B(a,r_0)$, then there is a constant $w_4\ge1$ independent of $a$
such that $w_4^{-1}\,d(x,y)\le|f_a(x)-f_a(y)|\le w_4\,d(x,y)$ for
any $x,y\in B(a,r_0)$.
\end{description}

\begin{lem}
                                  \label{ball}
If $0<r'\le r''$, then there is a number $N\in \mathbb{N}$
such that each ball of radius $r''$ can be covered by at most $N$
balls of radius $r'$. Moreover, $N\le
V_s((r'/2)+r'')/V_i(r'/2)$\,.
\end{lem}

\begin{proof} Let a ball $B(x,r'')$ be
given. Take a maximal system of points $x_1,\ldots,x_n$ from
$B(x,r'')$ such that the balls $B(x_j,r'/2)$ do not intersect each
other. Then the balls $B(x_j,r')$ cover $B(x,r'')$. On the other
hand, $V(x,(r'/2)+r'')\ge n V_i(r'/2)$, hence $n\le
V_s((r'/2)+r'')/V_i(r'/2)$.
\end{proof}

\subsection{Spaces and Kernels.}
Let $f$ be a measurable function on $X$; if $f\in L^p(X)$, $1\le
p\le\infty$, then $\|f\|_p$ denotes the norm of $f$ in $L^p(X)$,
otherwise we write $\|f\|_p=\infty$. Let $S$ be a bounded linear
operator from $L^p(X)$ to $L^q(X)$ with norm $\|S\|_{p,q}$. Such
an operator always has a kernel $K=K_S$ in the sense of
distributions; if $K\in L^1_{\rm loc}(X\times X)$, then $K$ is
called an \emph{integral kernel} of $S$. The operator $S$ with an
integral kernel $K_S$ is called an \emph{integral} operator if for
$f\in L^p(X)$ and for a.e. $x\in X$ we have
$K_S(x,\cdot)f(\cdot)\in L^1(X)$ (see e.g. \cite{HS}; note that we
consider only everywhere defined integral operators according to
the terminology of \cite{HS}). In virtue of the Closed Graph
Theorem, we have for an integral operator $S$ with the kernel $K$
\[
Sf(x)=\int_X\,K(x,y)f(y)\,dx\,\quad \text{for \,\,a.e.}\,\,\,x\,.
\]
Note that $S$ having an integral kernel is not necessary an
integral operator in the above sense: the simplest example is the
Fourier transform in $L^2(\mathbb{R}^\nu)$. Another example
related to the subject of the paper is the resolvent $R(\zeta)$ of
the free Hamiltonian $-\Delta$ in $L^2(\RR^\nu)$ for $\nu\ge 4$:
$R(\zeta)$ is not an integral operator in $L^2(\RR^\nu)$ but has
an integral kernel (the Green function). The
Gelfand--Dunford--Pettis Theorem gives a useful criterion for $S$
to be an integral operator; before we state this theorem we agree
on a following notation: If $r,s$ is another pair of numbers with
$1\le r,s\le \infty$ we denote $\|S\|_{r,s}=\sup\{\|Sf\|_s:\,f\in
L^p(X)\cap L^r(X), \|f\|_r\le1\}$ (the equality
$\|S\|_{r,s}=\infty$ is not excluded. Evidently, this definition
does not lead to contradiction in the case $p=r$, $q=s$. Now we
state the Gelfand--Dunford--Pettis Theorem in the form given
in~\cite[\S3.3]{KS}:

\begin{thm}\label{thma}
Let $S$ be a bounded operator from $L^p(X)$ to $L^\infty(X)$ and
$p<\infty$. Then $S$ is an integral operator and we have for its
kernel $K_S$
\[
\|S\|_{p,\infty}=\supess\limits_{x\in X}\|K_S(x,\cdot)\|_{p'}\quad
{ with}\quad p'=(1-p^{-1})^{-1}\,.
\]
In particular, if $S$ is a bounded operator from $L^p(X)$ to
$L^q(X)$ with $p<\infty$ and for some $r<\infty$ the condition
$\|S\|_{r,\infty}<\infty$ is satisfied, then $S$ has an integral
kernel $K_S$ and $\supess\limits_{x\in
X}\|K_S(x,\cdot)\|_{r'}<\infty$.
\end{thm}

If $K_1$ and $K_2$ are two integral kernels of $S$, then
$K_1(x,y)=K_2(x,y)$ a.e. in $X\times X$. If, in addition, $K_1$
and $K_2$ are {\it separately continuous} on $(X\times X)\setminus
D$, then $K_1=K_2$ everywhere on $(X\times X)\setminus D$.

An integral kernel $K(x,y)$ is called a {\it Carleman} kernel if
\[
\int_{X}|K(x,y)|^2\,dy<\infty\quad\, \text{for a.e. } x\in
X \,.
\]
A bounded operator on $\,L^2(X)$ having a Carleman kernel is
called also a \emph{Carleman operator}. It is clear that any
Carleman operator is an integral operator.

\begin{rem} By the definition from \cite{Sim}, Carleman
kernels $K$ obey the following additional condition
$\supess\limits_{x\in X}\|K(x,\cdot)\|_{2}<\infty$. We use the
definition of Carleman kernels from \cite{Kor}, which is wider
then that from \cite{Sim}.
\end{rem}

Fix $r>0$ and for each real $p$, $p\ge1$, introduce the space
$L^p_{\rm unif}(X)$ (\emph{uniformly local $L^p$-space}) by
\[
L^p_{\rm unif}(X)=\{f\in L^p_{\rm loc}(X)\,:\,\sup\limits_{x\in
X}\,\int_{B(x,r)}\,|f(x)|^p\,dx<\infty\}\,
\]
with the norm
\[
\|f\|_{p}^{(r)}=\bigg( \sup\limits_{x\in
X}\,\int_{B(x,r)}\,|f(x)|^p\,dx\bigg)^{1/p}\,.
\]
According to Lemma~\ref{ball}, the definition of $L^p_{\rm unif}$
is independent of $r$ and all the norms $\|\cdot\|_{p}^{(r)}$ with
$p$ fixed are mutually equivalent; we will denote
${\|\cdot\|_{p}^{(r_0)}}$ simply by ${\|\cdot\|_{p,{\rm unif}}}$. It
is clear that $L^p+L^\infty\subset L^p_{\rm unif}\subset L^p_{\rm
loc}$ and $L^p_{\rm unif}\subset L^q_{\rm unif}$, if $p\ge q$.

\begin{lem}
                         \label{exp}
Let $f\in L^1_{\rm unif}(X)$, $p\ge 1$, and $\omega>\theta_X$.
Then for each $a\in X$, the function
$g_a(x)=\exp(-\omega\,\,d(a,x)^p)f(x)$ belongs to $L^1(X)$  and
$\|g_a\|_1\le c\, \|f\|_{1,\,\, {\rm unif}}$, where the constant
$c$ depend on $\omega$ only. Moreover,
\[
\int_{d(a,x)\ge r}\,|g_a(x)|\,dx\,\to\,0\quad\text{as
$r\to\infty$}
\]
uniformly with respect to $a$ and to $f$ in the unit ball of
$L^1_{\rm unif}(X)$.
\end{lem}

\begin{proof}  Let $n\in\mathbb{N}$ be arbitrary,
then
\begin{multline*}
\int_{B(a,n)}|g_a(x)|\,dx=\sum_{k=1}^n\,\int_{B(a,k)\setminus
B(a,k-1)}|g_a(x)|\,dx\\
\le\|f\|_{1,\,\,\rm unif}\sum_{k=1}^n
N_k\exp(-\omega(k-1))\,,
\end{multline*}
where $N_k$ is the minimal number of balls of radius $r_0$
covering the ball $B(a,k)$. Using Lemma~\ref{ball} and the
estimate (V2), we get $N_k\le c''\exp(\theta_X\, k)$, where $c''$
is independent of $n$. Passing to the limit $n\to\infty$, we get
the estimate $\|g_a\|_1\le c\, \|f\|_{1,\,\, {\rm unif}}$.

Represent now $\omega$ in the form $\omega=\omega'+\omega''$,
where $\omega'>\theta_X$, $\omega''>0$. Then
\begin{multline*}
\int\limits_{d(a,x)\ge r}\,|g_a(x)|\,dx\,\le\,\exp(-\omega'' r^p)
\int\limits_X\exp(-\omega'd(a,x)^p)|f(x)|\,dx\\
\le \,c'\exp(-\omega'' r^p)\|f\|_{1,\,{\rm unif}}\,.
\end{multline*}
\end{proof}

\subsection{Self-adjoint operators.}\label{sec-sa}
Let $S$ be a self-adjoint operator in $L^2(X)$, not necessarily
bounded. We denote by $\spec S$ the spectrum of $S$ and by $\res
S$ the resolvent set $\CC\setminus\spec S$. For $\zeta\in \res S$
we denote by $R_S(\zeta)$ (or simply by $R(\zeta)$) the resolvent
of $S$: $R_S(\zeta)=(S-\zeta)^{-1}$. The kernel of $R(\zeta)$ in
the sense of distributions is called the {\it Green function} of
$S$ and will be denoted by $G_S(x,y;\zeta)$. For $\kappa>0$ and
$\zeta\in \res S$, $\Re\zeta<\inf\spec S$, we will consider the
power $R_S^\kappa(\zeta)$ of $R(\zeta)$ defined by
\begin{equation}
                      \label{fres}
R_S^\kappa(\zeta)=
\frac{1}{\Gamma(\kappa)}\int_0^{\infty}e^{-t(S-\zeta)}\,t^{\kappa-1}\,dt,
\end{equation}
where the integral is taken in the space of bounded operators in
$L^2(X)$ (it converges absolutely there). It is clear that for an
integer $\kappa$, Eq.~\eqref{fres} gives the usual power of
$R(\zeta)$. The (distributional) kernel for $R_S^\kappa(\zeta)$
will be denoted by $G_S^{(\kappa)}(x,y;\zeta)$. Note, that instead
of $S$ we will use for resolvents, propagators and their kernels
other subscripts identifying the operator $S$ and will omit these
subscripts, if confusion is excluded.

For numbers $p$ and $q$ with $1\le p\le q\le\infty$ we will
consider the following condition on the operator $S$ assuming $S$
is semibounded below:

\begin{description}
\item[(S${}_{pq}$)] there exist constants $B_{p,q}>\max(-\inf\spec
S,0)$ and $C_{p,q}>0$ such that for every $t>0$
\[
                     \label{Gri4}
\|e^{-tS}\|_{p,q}\le C_{p,q}\,t^{-\gamma}\exp(B_{p,q}\,t)\,,
\,\,\,\,
\text{where}\quad\,\gamma=\frac{1}{2}\nu(p^{-1}-q^{-1})\,.
\]
\end{description}
The proof of the following Theorem~\ref{thmb} is contained in the
proofs of the Theorems B.2.1--B.2.3 in \cite{Sim}. We include the
proof for sake of completeness.

\begin{thm}\label{thmb} Let $S$ be a self-adjoint semibounded
below operator in $L^2(X)$ obeying the condition $({\rm
S}{}_{pq})$ from Subection~\ref{sec-sa} for some $p$ and $q$ with $1\le p\le q\le\infty$ and
let $\kappa>0$ with $p^{-1}-q^{-1}<2\kappa/\nu$ be given.
Then the following assertions are true:
\begin{enumerate}
\item[$(1)$] $\|R^\kappa(\zeta)\|_{p,q}<\infty$, if one of the
following conditions is satisfied:
\begin{itemize}
\item[(a)] $\Re\zeta<-B_{p,\infty}$;
\item[(b)] $\kappa$ is an integer and
$p\le2\le q$.
\end{itemize}
Moreover, $\|R^\kappa(\zeta)\|_{p,q}\to 0$ as
$\Re\zeta\to-\infty$.

\item[$(2)$] Assume additionally that $p\le2\le q$. Take a Borel
function $f$ on $\spec{S}$ satisfying for $\xi\in\spec S$ the
estimate $\big|\,f(\xi)\big|\le b\big(|\xi|+1\big)^{-\kappa}$ with some $b>0$. Then
$\|f(S)\|_{p,q}<Cb$ where $C>0$ is independent of $b$. Suppose
$q=\infty$, then $f(S)$ is an integral operator, if
$\kappa>\nu/2p$. Moreover, in the last case $f(S)$ is a Carleman
operator if $p=2$ and has has an integral kernel $F(x,y)$ bounded
by the constant $Cb$ if $p=1$.

\item[$(3)$] Suppose $q=\infty$ and take $\zeta\in \res S$. Then
$R^\kappa(\zeta)$ has an integral kernel if one of the following
conditions is satisfied: $\rm(a)$ $\Re\zeta<-B_{p,\infty}$;
$\rm(b)$ $\kappa$ is an integer and $p\le2$. Moreover, in both
cases, $R^\kappa(\zeta)$ is a Carleman operator if $p=2$, and  the
integral kernel of $R^\kappa(\zeta)$ is bounded, if $p=1$.
\end{enumerate}

\end{thm}

\begin{proof}

According to \eqref{fres},
\[
\|R^\kappa(\zeta)\|_{p,q}\le\frac{1}{\Gamma(\kappa)}
\int_0^{\infty}\|e^{-tS}\|_{p,q}e^{t\Re\zeta}\,t^{\kappa-1}\,dt,
\]
if $\Re\zeta<-B_{p,q}$. Therefore, in this case (1a) is proven and
$\|R^\kappa(\zeta)\|_{p,q}\to 0$ as $\Re\zeta\to-\infty$. Before
completing the proof of the item (1) we prove the item (2).

Fix $E$, $E<B_{p,q}$, and let $g(\xi)=(\xi-E)^{\kappa}f(\xi)$.
Represent $\kappa$ as the sum $\kappa=\kappa'+\kappa''$ such that
$p^{-1}-2^{-1}<2\kappa'/\nu$, $2^{-1}-q^{-1}<2\kappa''/\nu$, then
by (1a) we have $b_1:=\|R^{\kappa'}(E)\|_{p,2}<\infty$ and
$b_2:=\|R^{\kappa''}(E)\|_{2,q}<\infty$. Moreover, $|g(\xi)|\le
b_3<\infty$ for all $\xi\in\spec S$. Since $f(S)=
R^{\kappa''}(E)g(S)R^{\kappa'}(E)$, we get $\|f(S)\|_{p,q}\le bC$
with $b=b_1b_2b_3$. The last statements of the item (2) follow
immediately from Theorem~\ref{thma}. The subitem (1b) follows
easily from (2); the item (3) is the consequence of
Theorem~\ref{thma} and items (1), (2).
\end{proof}

\begin{corol} \label{corol4} Let the operator $S$ satisfy
the conditions of Theorem~\ref{thmb}. Then the following
assertions are true:
\begin{enumerate}
\item[$(1)$] Let $0\le p^{-1}-q^{-1}<2/\nu$ and numbers $r,s$ be
taken such that $1\le r,s\le\infty$, $r^{-1}=p^{-1}-s^{-1}$. If
$W\in L^s(X)$, then $\|R(E)W\|_{r,q}<\infty$ for $E<0$ with
sufficiently large $|E|$; moreover $\|R(E)W\|_{r,q}\to 0$ as
$E\to-\infty$.

\item[$(2)$] Let in addition $q=\infty$. Then for any $W\in
L^p(X)$ we have

\begin{enumerate}
\item[$(2{\rm a})$] $\|R(E)W\|_{\infty,\infty}<\infty$ for $E<0$
with sufficiently large $|E|$ and $\|R(E)W\|_{\infty,\infty}\to 0$
as $E\,\to-\infty\,$;

\item[$(2{\rm b})$] $\|\,|W|^{1/2}R(E)|W|^{1/2}\|_{2,2}<\infty$
for $E<0$ with sufficiently large $|E|$ and
$\|\,|W|^{1/2}R(E)|W|^{1/2}\|_{2,2}\to0$ as $E\to-\infty$.
\end{enumerate}
\end{enumerate}
\end{corol}

\begin{proof} (1) Since $W$ is a continuous mapping from
$L^r$ to $L^p$, the proof follows from the item (1) of the
theorem.

\noindent (2a) This item is a particular case of (1).

\noindent(2b) It follows from (2a) by duality
$\||W|\,R(E)\,\|_{1,1}\to0$ as $E\to-\infty$, therefore the item
(2b) follows from the Stein interpolation theorem (see the
approach (2a) to the proof of Formula~(A26) in \cite{Sim}).
\end{proof}

\begin{rem}
We emphasize that the item (3) of Theorem~\ref{thmb} can be
considerably refined for functions of Schr\"odinger operators in
the Euclidian spaces, see e.g.~\cite{BGK,GerK}.
\end{rem}

For our purpose, a class of Carleman operators $S$ in $L^2(X)$ is
important; this class consists of operators with integral kernels
$K$ having the following continuity conditions:

\begin{description}
\item[(C1)]
for every $f\in L^2(X)$ the function
$g_f(x)= \displaystyle\int_{X}K(x,y)f(y)\,dy$ is continuous;
\item[(C2)]
the function $X\ni x\mapsto\displaystyle\int_{X}|K(x,y)|^2\,dy$ is continuous.
\end{description}

\begin{rem} In virtue of $\rm(C1)$ the image of an operator $S$ with the
corresponding kernel $K$ consists of continuous functions.
Moreover, $S$ is a continuous mapping from $L^2(X)$ to the space
$C(X)$ endowed with the topology of uniform convergence on
compact sets. Note that  the inclusion
$S\big(L^2(X)\big)\subset C(X)$ alone does not imply the continuity of the functions
in $\rm(C1)$, these need only be continuous after a modification
on a set of measure zero.
\end{rem}

\begin{prop} \label{corol5}  If a kernel $K$ fulfils the conditions
$\rm(C1)$ and $\rm(C2)$, then the mapping $F:\,x\mapsto
K(x,\cdot)$ from $X$ to the Hilbert space $L^2(X)$ is continuous.
\end{prop}

\begin{proof} The condition (C1) shows that $F$ is continuous with
respect to the weak topology of $L^2(X)$, and (C2) implies that
$x\mapsto\|F(x)\|$ is continuous.
\end{proof}

Using the proofs of Lemmas~B.7.8 and B.7.9 from  \cite{Sim}, we
obtain easy the following theorem:

\begin{thm}\label{thmc} \textup{(1)} Let $Q$, $S$, and $T$ be
bounded operators in $L^2(X)$ such that $S$ and $T$ have Carleman
kernels with the properties \textup{(C1)} and \textup{(C2)} above. Then
$S^*QT$ is a Carleman operator with a continuous kernel in
$X\times X$.

\medskip

\textup{(2)} Let $S$ be a self-adjoint operator in $L^2(X)$ and
$f$ be a Borel function on $\spec S$ such that for all
$\xi\in\spec S$ there holds
$\big|f(\xi)\big|\le b\,\big(|\xi|+1\big)^{-2\kappa}$
with $b>0$, $\kappa>0$. If for some $\zeta\in \res S$ the operator
$R_S^\kappa(\zeta)$ has a Carleman kernel with properties $\rm
(C1)$ and $\rm (C2)$, then $f(S)$ is a Carleman operator and its
kernel $F(x,y)$ is continuous in $X\times X$. Moreover, if
$\|R_S^\kappa(\zeta)\|_{2,\infty}\le c$, then $|F(x,y)|\le b\,c^2$
for all $x,y\in X$.
\end{thm}

\subsection{Schr\"odinger operators and related kernels.}
We denote by $H_0$ the Laplace--Beltrami operator on $X$,
$H_0=-\Delta$ (the Schr\"odinger operator of a free charged
particle on $X$). The corresponding resolvent, the Green function
and the integral kernel of the Schr\"odinger semigroup (heat
kernel) $e^{-tH_0}$ are denoted by $R_0(\zeta)$,
$G_0(x,y\,;\zeta)$, and $P_0(x,y\,;t)$, respectively. Let
$A=\sum_{j=1}^\nu A_j\,dx^j$ be a 1-form on $X$, for simplicity we
suppose here $A_j\in C^\infty(X)$. The functions $A_j$ can be
considered as the components of the vector potential of a {\it
magnetic field} on $X$. On the other hand, $A$ defines a
connection $\nabla_A$ in the trivial line bundle $X\times
\mathbb{C}$, $\nabla_A u=du+iuA$; by
$-\Delta_A=\nabla_A^*\nabla_A$ we denote the corresponding Bochner
Laplacian. The operator $H_A=-\Delta_A$ is essentially
self-adjoint on $C^\infty_0(X)$. In addition, we consider a scalar
potential $U$ of an electric field on $X$, which is a real-valued
measurable function, $U\in L^2_{\rm loc}(X)$; if $H_A+U$ is
essentially self-adjoint on $C_0^\infty(X)$, then its closure (the
{\it magnetic Schr\"odinger operator}) is denoted by $H_{A,U}$.
The corresponding resolvent, Green function, and the heat kernel
will be denoted by $R_{A,U}$, $G_{A,U}$, and $P_{A,U}$,
respectively.

For real valued functions $U$ on $X$ we denote as usual
$U_+:=\max(U,0)$, $U_-:=\max(-U,0)\equiv U_+-U$. The following result of
M.~Shubin plays one of the crucial part below (see
\cite[Theorem~1.1]{Shu1}):

\begin{thm}\label{thmd}   Let $U$ be a real-valued function
on $X$ such that $U_+\in L^2_{\rm loc}(X)$ and $U_-\in L^p_{\rm
loc}(X)$ with $p=\nu/2$ if $\nu\ge5$, $p>2$ if $\nu=4$, and $p=2$
if $\nu\le3$. If $H_A+U$ is semi-bounded below on $C_0^\infty(X)$,
then $H_A+U$ is essentially self-adjoint on $C_0^\infty(X)$.
%
\end{thm}

The properties of $P_0(x,y\,;t)$  we need below are
presented in the following theorem, see \cite[Formula 3.14]{Gri1}:

\begin{thm}\label{thme}
The function $P_0(x,y\,;t)$ is of
class $C^\infty$ on $X\times X\times(0,\infty)$ and
\begin{equation}
                      \label{Gri1}
0\le P_0(x,y\,;t)\le
\frac{C_P}{\min(t^{\nu/2},1)}\,\left(1+\frac{d(x,y)^2}{t}\right)^{\tfrac{\nu}{2}+1}\,
\exp\left(-\frac{d(x,y)^2}{4t}-\lambda t\right)\,,
\end{equation}
where $C_P>0$, $\lambda=\inf\,\spec{H_0}$. Moreover,
\begin{equation}
                      \label{Gri2}
\sup\limits_{x,t}\,\int_X P\,{}_0(x,y\,;\,t)\,dy\le 1\,.
\end{equation}
\end{thm}

\subsection{Kato's inequality.}
We recall that a linear operator $S\,: L^p(X)\rightarrow L^q(X)$
is said to be {\it positive in the sense of the point-wise order}
or {\it positivity preserving}, if $Sf(x)\ge0$ a.e for every $f\in
L^p(X)$ with $f(x)\ge0$ a.e.; such an operator is bounded
\cite{Vul}. A positive operator $S$ {\it dominates} a linear
operator $T\,: L^p(X)\rightarrow L^q(X)$ in the sense of the
point-wise order, if for all $f\in L^p(X)$ we have $|Tf(x)|\le
S|f|(x)$ a.e. If $T$ is dominated by a positive operator, then $T$
is bounded and there is a positive operator $|T|\,:
L^p(X)\rightarrow L^q(X)$ with the following properties:
\begin{itemize}
\item[(1)] $T$ is dominated by $|T|$,
\item[(2)] if $S$ is another positive operator
which dominates $T$, then $S-|T|$ is positive preserving (in
symbols: $|T|\le S$ {\it in the sense of the point-wise order});
it is clear that in this case $\|T\|_{p,q}\le \|S\|_{p,q}$.
\end{itemize}
Moreover, if $q=\infty$, then we have for integral kernels
$|K_T(x,y)|\le K_S(x,y)$ a.e.

The main tool to extend results
obtained for a Schr\"odinger operator without magnetic fields to
that with a nontrivial magnetic field is the following theorem
which combines \cite[Theorem~5.7]{BMS} and \cite[Theorem 2.15]{HSU}.

\begin{thm}\label{thmf}
Let $U$ satisfy the condition of Theorem~\ref{thmd} and let
$H_A+U$ be semibounded below on $C^\infty_0(X)$. Then the
following assertions are true.
\begin{enumerate}
\item[$(1)$] Semigroup dominations: For every $t$, $t>0$, we have
$|e^{-tH_{A,U}}|\le e^{-tH_{0,U}}$ in the sense of the point-wise
order in $L^2(X)$; hence, $|P_{A,U}(x,y;t)|\le P_{0,U}(x,y;t)$ for
a.e. $x,y\in X$.
\item[$(2)$] Resolvent dominations: For every $E$, $E<0$, with sufficiently
large $|E|$, we have $|R_{A,U}(E)|\le R_{0,U}(E)$ in the sense of
the point-wise order in $L^2(X)$; hence, $|G_{A,U}(x,y;E)|\le
G_{0,U}(x,y;E)$ for a.e. $x,y\in X$.
\end{enumerate}
\end{thm}

\subsection{Admissible potentials, convenient kernels}\label{cond-pot}

The main results of the paper require some properties of
considered potentials and kernels. We call a potential $U$
\emph{admissible}, if  $U_+\in L^{p_0}_{\rm loc}(X)$ and
$U_-\in \sum_{i=1}^n L^{p\,{}_i}(X)$, where
$2\le p_i\le\infty$ if $\nu\le3$, $\nu/2<p_i\le \infty$ if
$\nu\ge4$ ($0\le i\le n$) (we stress that $p_i$ depend on $U$).
The class of all admissible potentials will be denoted by
$\mathcal{P}(X)$. It is clear that $\mathcal{P}(X)$ is a saturated
cone in the space of all measurable real valued functions
$L^0_\mathbb{R}(X)$ on $X$, i.e. if $U_1,U_2\in\mathcal{P}(X)$,
then
\begin{itemize}
\item $\lambda_1,\lambda_2\in
\mathbb{R}_+$ implies $\lambda_1 U_1+\lambda_2 U_2\in
\mathcal{P}(X)$;
\item $V\in L^0_\mathbb{R}(X)$,
and $U_1\le V \le U_2$ implies $V\in\mathcal{P}(X)$.
\end{itemize}
We show in Section~3 that  $H_A+U$ is essentially self-adjoint and
semi-bounded below on $C_0^\infty(X)$ if $U\in\mathcal{P}(X)$.

To use the Lippmann--Schwinger equation we need some restriction
on the integral kernels which control the behavior of the kernels
near the diagonal and on the infinity. The norm estimates of the
Green functions from Theorem~\ref{thmb} show a usefulness of the
following classes of kernels. Let $0\le\alpha<\nu$, $1\le
p\le\infty$. We denote by $ \mathcal{K}(\alpha,p)$ the class of
all measurable functions $K$ {\it everywhere defined} on $(X\times
X)\setminus D$ if $\alpha>0$ and on $X\times X$ if $\alpha=0$, and
obeying the conditions

\begin{description}
\item[(L1)] for a constant $c=c(K)>0$ there holds $|K(x,y)|\le c
\max(1,d(x,y)^{-\alpha})$ for {\it all} $(x,y)\in (X\times
X)\setminus D$ if $\alpha>0$ and for {\it all} $x,\,y\in X$
otherwise; \item[(L2)] for every $r>0$ (or, which is the same, for
all sufficiently small $r>0$) there holds
\begin{multline*}
\lfloor K\rfloor_{p,r}:=\displaystyle\max\Big(\,\supess\limits_{x\in X}
\big\|\chi_{X\setminus
B(x,r)}\,K(x,\cdot)\big\|_p\,,\\
\supess\limits_{y\in X}
\big\|\chi_{X\setminus B(y,r)}\,K(\cdot,y)\big\|_p\,\Big) <\infty\,,
\end{multline*}
where $\chi_A$ denotes the characteristic function of
$A\subset X$.
\end{description}

By $\mathcal{K}_0(\alpha,p)$ we will denote the subclass of
$\mathcal{K}(\alpha,p)$ consisting of all functions $K$ from
$\mathcal{K}(\alpha,p)$ obeying the condition

\begin{description}
\item[(L3)] $\lim\limits_{r\to\infty}\lfloor K\rfloor_{p,r}=0$.
\end{description}

\noindent Below we list the simplest properties of the classes
$\mathcal{K}(\alpha,p)$ and $\mathcal{K}_0(\alpha,p)$ which are
needed below.

\begin{description}

\item[(K1)] If $\alpha>0$, then the condition (L1) is equivalent
to each of the following ones:

\begin{description}

\item[(L1a)] for a constants $r>0$ and $c>0$ there holds:
$|K(x,y)|\le c\,d(x,y)^{-\alpha}$ if $0<d(x,y)<r$ and $|K(x,y)|\le
c$ if $d(x,y)\ge r$;

\item[(L1b)] for every $r>0$ there is a constants $c>0$ such that:
$|K(x,y)|\le c\,d(x,y)^{-\alpha}$ if $0<d(x,y)<r$ and $|K(x,y)|\le
c$ if $d(x,y)\ge r$.
\end{description}
For $\alpha=0$ the condition $0<d(x,y)<r$ must be
replaced by $d(x,y)<r$.

\item[(K2)] $\mathcal{K}(\alpha_1,p)\subset
\mathcal{K}(\alpha_2,p)$ and $\mathcal{K}_0(\alpha_1,p)\subset
\mathcal{K}_0(\alpha_2,p)$, if  $0\le\alpha_1\le\alpha_2<\nu$ and
$1\le p\le\infty$.

\item[(K3)] If $K$ satisfies (L1), then
$K\in\mathcal{K}(\alpha,\infty)$. In particular,
$\mathcal{K}(\alpha,p)\subset \mathcal{K}(\alpha,\infty)$ for all
$p\ge1$ and $\alpha$, $0\le\alpha<\nu$. Therefore
$\mathcal{K}(\alpha,p)\subset \mathcal{K}(\alpha,q)$ if $p\le q\le
\infty$, and $\mathcal{K}_0(\alpha,p)\subset
\mathcal{K}_0(\alpha,q)$ if $p\le q< \infty$.

\item[(K4)] All the classes $\mathcal{L}=\mathcal{K}(\alpha,p)$ or
$\mathcal{L}=\mathcal{K}_0(\alpha,p)$ are order ideals in the
space of measurable functions $L^0(X\times X)$, i.e. $\mathcal{L}$
is a linear subspace in $L^0(X\times X)$ with the property: If
$K\in \mathcal{L}$, $L\in L^0(X\times X)$ and $|L(x,y)|\le
|K(x,y)|$ for all $(x,y)$, then $L\in \mathcal{L}$\,.
\end{description}
{}From Lemma~\ref{sing} below (Section~\ref{sec2}) we get
obviously the following property

\begin{description}
\item[(K5)] $\mathcal{K}(\alpha,p)\subset L^1_{\rm loc}(X\times
X)$ for all $p$ and $\alpha$ with $1\le p\le\infty$,
$0\le\alpha<\nu$.
\end{description}

\noindent The next lemma delivers an important class of functions
belonging to $\mathcal{K}_0(\alpha,p)$.

\begin{lem}\label{K0}
Let $P(x,y;t)$ be a measurable function on $X\times X\times
(0,\infty)$ such that for some constants $c_j$, $c_j>0$, $j=1,2,3$,
we have the estimate
\begin{equation}
                       \label{K0est}
\big|P(x,y;t)\big|\le c_1t^{-\nu/2}\exp\Big(c_2 t-c_3\dfrac{d(x,y)^2}{t}\Big)\,.
\end{equation}
Let $\alpha=0$ if $\nu=1$, $\alpha$ be an arbitrary number from
the interval $(0,\nu)$ if $\nu=2$, and $\alpha=\nu-2$ if
$\nu\ge3$. Let $\gamma\ge 0$; for any $\zeta\in\mathbb{C}$,
$\Re\zeta<-c_2$, define the following kernel $K_\gamma(x,y;\zeta)$
by the Laplace transform
\[
K_\gamma(x,y;\zeta)=\int_0^\infty\,t^\gamma\,P(x,y;t)\,e^{t\zeta}\,dt\,.
\]
Then $K_\gamma$ belongs to all the classes
$\mathcal{K}_0(\alpha,p)$ with $1\le p\le\infty$ for all $\zeta$
with $\Re\zeta$ sufficiently close to $-\infty$.
\end{lem}

\begin{proof} Since the kernel $t^\gamma P(x,y;t)$ admits the
estimate of the type (\ref{K0est}), it sufficient to consider the
case $\gamma=0$ only. It is well known that for any fixed $c$,
$c>0$, the function $K_0(x,E)$ defined for $x>0$, $E<0$ as the
integral
\[
K_0(x,E):=\int_0^\infty
t^{-\nu/2}\exp(Et-cx^2t^{-1})\,dt\,,
\]
has the following asymptotic properties:
\begin{itemize}
\item[(1)] for fixed $E<0$ there holds $K_0(x,E)=O(h(x))$ as
$x\to0$, where
\[
h(x)=\begin{cases}
    x^{-\nu+2}\,, & \text{if $\nu>2$}, \\
    |\log x| \,, & \text{if $\nu=2$}, \\
    1\,, & \text{if $\nu=1$}.
  \end{cases}
\]
\item[(2)] For every $\delta>0$ there exist
$c'_\delta,c''_\delta>0$ such that for $|x|\ge\delta$ we have
$K_0(x,E)\le c'_\delta\exp(-c''_\delta x)$; here $c'_\delta$ is
independent of $E$ with $E\le-1$ whereas $c''_\delta\to+\infty$ as
$E\to-\infty$.
\end{itemize}
Now the property (L1) for $K$ follows from (1) and (L3) from (2)
(if $p<\infty$ we use additionally Lemma~\ref{exp}).
\end{proof}

\section{Auxiliary results concerning convenient kernels}\label{sec2}
\begin{lem}\label{sing}
\begin{enumerate}
\item[$(1)$]
Let $\alpha\in\mathbb{R}$, $\alpha<\nu$, and $a\in X$. Then for
every $x\in X$ and $r>0$ there holds
\begin{equation}
                          \label{lem1}
J_1(x):=
\int_{B(a,r)}\,d(x,y)^{-\alpha}\,dy<\infty\,.
\end{equation}
Moreover, there exists a constant $\tilde c_\alpha$ depending only
on $\alpha$, such that if $r\le r_0/3$, then
\[
J_1(x)\le
  \begin{cases}
\tilde c_\alpha r^{\nu-\alpha}\,,   & \text{if $\alpha>0$}, \\
\tilde c_\alpha\,r^{\nu}\,(r+d(a,x))^{-\alpha}  &
\text{otherwise}.
  \end{cases}
\]
\item[$(2)$] Let $0<\alpha_1,\,\alpha_2<\nu$ and
$\beta=\alpha_1+\alpha_2-\nu$. Then there is a constant $c>0$ such
that for any $a\in X$ and any $r$, $0<r<r_0$, we have for $x,y\in
B(a,r)$, $x\ne y$:
\end{enumerate}
\[
J_2(x,y):= \int\limits_{B(a,r)}\,d(x,z)^{-\alpha_1}d(y,z)^{-\alpha_2}\,dz
\le\begin{cases}
    c\,d(x,y)^{-\beta}\,, & \text{if $\beta>0$}, \\
    c\,\big(|\log d(x,y)|+1\big)\,, & \text{if $\beta=0$}, \\
    c\,, & \text{otherwise.}
  \end{cases}
\]
\end{lem}

\begin{proof} (1) According to (V5) we can choose $c'>0$ in such a
way that $V_s(r)\le c'r^\nu$ if $r\le r_0$.

Let $\alpha\le0$, then $d(x,y)\le r+d(a,x)$ for $y\in B(a,r)$,
therefore
\[
\int_{B(a,r)}\,(d(x,y))^{-\alpha}\,dy\le
(r+d(a,x))^{-\alpha}V_s(r)\,,
\]
in particular, for $r\le r_0$
\begin{equation}
                 \label{Ap2aa}
J_1(x)\le c'r^\nu (r+d(a,x))^{-\alpha}\,.
\end{equation}

Let now $\alpha>0$. Suppose firstly $d(a,x)\ge 2r$. Then for $y\in
B(a,r)$ we have $d(x,y)\ge d(a,x)-d(a,y)\ge r$. Therefore
\[
\int_{B(a,r)}\,(d(x,y))^{-\alpha}\,dy\,\le\,V_s(r)r^{-\alpha}\,.
\]
In particular, for $r\le r_0$
\begin{equation}
                 \label{Ap2a}
J_1(x)\le\,c'\,r^{\nu-\alpha}\,,
\end{equation}
Now suppose $d(a,x)<2r$. Then for $y\in B(a,r)$ we have $d(x,y)\le
d(a,x)+d(a,y)<3r$. Therefore
\[
\int_{B(a,r)}\,(d(x,y))^{-\alpha}\,dy\le
\int_{B(x,3r)}\,(d(x,y))^{-\alpha}\,dy=
\int_0^{3r}\,{\rho^{-\alpha}}\,dV(x,\rho)\,,
\]
where the integral in the right-hand side is Stieltjes with
respect to the increasing function $\rho\rightarrow V(x,\rho)$.
Using the integration by part we get:
\[
\int_{B(x,3r)}\,(d(x,y))^{-\alpha}\,dy=
V(x,3r)(3r)^{-\alpha}+
\alpha\int_0^{3r}\,\frac{V(x,\rho)}{\rho^{\alpha+1}}\,d\rho\,<\infty
\]
since $V(x,\rho)\le c''\rho^\nu$ by (V5). In particular, if $r\le
r_0/3$, then
\begin{equation}
                        \label{Ap5}
J_1(x) \le \frac{3^{\nu-\alpha}\nu
c'}{\nu-\alpha}\,r^{\nu-\alpha}\,\,.
\end{equation}
Now the result follows from (\ref{Ap2aa}), (\ref{Ap2a}), and
(\ref{Ap5}).

(2) Since property $(2)$ is local, it follows from (V3) and (V6)
that we can restrict ourselves by the proof for the case $X= \mathbb{R}^\nu$,
$y=0$. Then (2) follows from the inequality $|x-a|< r$ and the
following assertion:

{\it Let $\alpha_1,\alpha_2\in\mathbb{R}$ and
$\alpha_1,\alpha_2<\nu$, then for any $a\in \mathbb{R}^\nu$,
$r>0$, and $x\in B(a,r)$, $x\ne0$, there holds
\begin{multline}
                         \label{sing2}
I(x):=\int\limits_{B(a,r)}\frac{dz}{|x-z|^{\alpha_1}
|z|^{\alpha_2}}\\{}\le
  \begin{cases}
\dfrac{c'}{|x|^{\alpha_1+\alpha_2-\nu}}+
\dfrac{c''}{(|a|+r)^{\alpha_1+\alpha_2-\nu}},  & \text{if $\alpha_1+\alpha_2\ne\nu$}, \\
\displaystyle
 c'\log\frac{|a|+r}{|x|}+c'',
 & \text{otherwise,}
  \end{cases}
\end{multline}
where the constants $c'$ and $c''$ are positive and depend only on
$\alpha_1$ and $\alpha_2$.}

We start the proof of this assertion with the change of variables
$z=|x\,|u$ in the integral \eqref{sing2}; the result is
\[
I(x)=|x\,|^{\nu-\alpha_1-\alpha_2}
\int\limits_{B\,\big(\tfrac{a}{|x\,|},\tfrac{r}{|x\,|}\big)}
\,|e_x-u|^{-\alpha_1}|u|^{-\alpha_2}\,du\,,
\]
where $e_x=x/|x|$. Let $B=B\Big(0,\dfrac{|a|+r}{|x\,|}\Big)$, then
\[
I(x)\le |x\,|^{\nu-\alpha_1-\alpha_2}
\int_{B}\,|e_x-u|^{-\alpha_1}|u|^{-\alpha_2}\,du\,.
\]
Denote $B\,{}'=B(0,2)$ and
\[
c:=\int_{B\,{}'}\,|e_x-u|^{-\alpha_1}|u|^{-\alpha_2}\,du\,
\]
($c$ is independent of $x$ in virtue of the rotational symmetry
consideration). Since $|u|-1\le |e_x-u|\le1+|u|$, we have
$2^{-1}|u|\le |e_x-u|\le2|u|$ if $|u|\ge 2$; hence,
$|e_x-u|^{-\alpha_1}\le 2^{|\alpha_1|}|u|^{-\alpha_1}$ for such
values of $u$. Therefore
\begin{multline*}
I(x)\le|x\,|^{\nu-\alpha_1-\alpha_2}\bigg(c+2^{|\alpha_1|}\int_{B\setminus
B\,{}'}\,|u|^{-\alpha_1-\alpha_2}\,du\bigg)={}\\
|x\,|^{\nu-\alpha_1-\alpha_2}\bigg(c+2^{|\alpha_1|}s_\nu\int_2^{\tfrac{|a|+r}{|x\,|}}\,
\rho^{\nu-1-\alpha_1-\alpha_2}\,d\rho\bigg)\,,
\end{multline*}
where $s_\nu$ is the area of the unit sphere in $\mathbb{R}^\nu$.
Calculating the integral, we get the result.
\end{proof}

The following Lemma plays the main part in the article. Below we
denote as usual $p\,{}'=\dfrac{p}{p-1}$ for $1\le p\le \infty$.

\begin{lem}
           \label{lemma3}
Take $K_j\in \mathcal{K}(\alpha_j,q_j)$, $j=1,2$, and let
$W=\sum\limits_{k=1}^n\,W_k$, where $W_k\in L^{p_k}(X)$. Denote
$\displaystyle p\,{}_{\rm min}=\min_{1\le k\le n}\,p_k$,
$\displaystyle p\,{}_{\rm max}=\max_{1\le k\le n}\,p_k$ and
suppose that the following conditions are satisfied:

\begin{itemize}

\item[$(\rm a)$] $\displaystyle \frac{1}{p\,{}_{\rm
max}}+\frac{1}{q_1}+\frac{1}{q_2}=1$;

\item[$(\rm b)$] $p\,{}_{\rm min}\ge1$, if $\alpha_1=\alpha_2=0$,
and $p\,{}_{\rm min}>\nu/(\nu-\max(\alpha_1,\alpha_2))$ otherwise.
\end{itemize}
Then for the function $F(x,y,z)= K_1(x,z)W(z)K_2(z,y)$, the
following assertions are true.


\begin{enumerate}

\item[$(1)$] $F(x,y,\cdot)\in L^1(X)$ for $x\ne y$; therefore the
function $J$,
\[
J(x,y)=\int_X F(x,y,z)\,dz\,,
\]
is well-defined on $(X\times X)\setminus D$.

\item[$(2)$] Denote
$\alpha:=\max\big(0,\,\alpha_1+\alpha_2-\dfrac{\nu}{p\,{}'_1}\,,
\ldots,\, \alpha_1+\alpha_2-\dfrac{\nu}{p\,{}'_n} \big)$. Then
$J\in\mathcal{K}(\alpha,\infty)$ if
$p\,{}'_k(\alpha_1+\alpha_2)\ne\nu$ for all $k$, $k=1\,\ldots\,n$.
Otherwise $J\in\mathcal{K}(\alpha,\infty)$ if $\alpha>0$, and
$J\in\mathcal{K}(\beta,\infty)$ with arbitrary $\beta>0$, if
$\alpha\le 0$ $($we assume here $\infty\cdot0=0)$.


\item[$(3)$] Let $p\,{}_{\rm max}<\infty$ or at least one of the
functions $K_j$ $(j=1,2)$ belong to $\mathcal{K}_0(\alpha_j,q_j)$
with $\alpha_j$ and $q_j$ obeying the conditions $(\rm a)$ and
$(\rm b)$. Then the function $J$ has the continuity properties
listed below:

\begin{enumerate}

\item[$\rm(3a)$] if $K_1(\cdot,z)$ is continuous in
$X\setminus\{z\}$ for a.e. $z\in X$, then $J(\cdot,y)$ is
continuous in $X\setminus\{y\}$ $\rm for\,\,\,\,all$ $y\in X$;

\item[$\rm(3b)$]  if $K_2(z,\cdot)$ is continuous in
$X\setminus\{z\}$ for a.e. $z\in X$, then $J(x,\cdot)$ is
continuous in $X\setminus\{x\}$ $\rm for\,\,\,\,all$ $x\in X$;

\item[$\rm(3c)$]  if $K_1(\cdot,z)$ and $K_2(z,\cdot)$ are
continuous in $X\setminus\{z\}$ for a.e. $z\in X$, then $J$ is
continuous in $(X\times X)\setminus D$.

\end{enumerate}

\item[$(4)$] Let $\alpha_1+\alpha_2<\nu$. If
$\alpha_1+\alpha_2\ne0$, assume additionally that $W_k\in
L^q_\mathrm{loc}(X)$ for some $q>\nu/(\nu-\alpha_1 -\alpha_2)$ and
all $k$, $k=1\,,\ldots,\,n$. Then $F(x,y,\cdot)\in L^1(X)$ for all
$x,y\in X$, so that $J$ is well-defined on $X\times X$. Moreover,
if $p\,{}_{\rm max}<\infty$ or at least one of the conditions
$K_j\in\mathcal{K}_0(\alpha_j,q_j)$ $(j=1,2)$ is satisfied, then
the following continuity properties take place:

\begin{enumerate}
\item[$\rm(4a)$] if $K_1(\cdot,z)$ is continuous in
$X\setminus\{z\}$ for a.e. $z\in X$, then $J(\cdot,y)$ is
continuous in $X$ $\rm for\,\,\,\,all$ $y\in X$;

\item[$\rm(4b)$] If $K_2(z,\cdot)$ is continuous in
$X\setminus\{z\}$ for a.e. $z\in X$, then $J(x,\cdot)$ is
continuous in $X$ $\rm for\,\,\,\,all$ $x\in X$;

\item[$\rm(4c)$] If $K_1(\cdot,z)$ and $K_2(z,\cdot)$ are
continuous in $X\setminus\{z\}$ for a.e. $z\in X$, then then $J$
is continuous in $X\times X$.
\end{enumerate}
\end{enumerate}
\end{lem}

\begin{proof} First of all we conclude from the property (K3)
(see Section~\ref{cond-pot}) that for every $k$,
$k=1\,,\ldots\,,n$, there are $q_j^{(k)}$ such that
$K_j\in\mathcal{K}_j(\alpha_j,\,q_j^{(k)})$ and the following
properties are satisfied:

\begin{itemize}

\item[(${\rm a}_k$)] $\displaystyle
\frac{1}{p\,{}_{k}}+\frac{1}{q^{(k)}_1}+\frac{1}{q^{(k)}_2}=1$\,.
\end{itemize}
Moreover, it is clear that for all $k$

\begin{itemize}
\item[(${\rm b}_k$)] $p\,{}_{k}\ge1$, if $\alpha_1=\alpha_2=0$,
and $p\,{}_{k}>\nu/(\nu-\max(\alpha_1,\alpha_2))$ otherwise\,.
\end{itemize}
Therefore, to prove the properties (1), (3) and (4) we can suppose
$n=1$ since $J$ is additive with respect to $W$. This is true for
the property (2) as well, it sufficient to take into consideration
(K2) from Section~\ref{cond-pot}. Hence, further we consider the
case $n=1$ only.

\medskip

(1) Fix $x,y\in X$ such that $x\ne y$ and take $\eta$,
$0<\eta<d(x,y)/2$. In the ball $B(x,\eta)$, we estimate
$|F(x,y,z)|\le c d(x,z)^{-\alpha_1}|W(z)|$; therefore if
$\alpha_1=0$, the inclusion $F(x,y,\cdot)\in L^1(B(x,\eta))$ is
obvious. If $\alpha_1>0$, the inequality
$p>\nu/(\nu-\max(\alpha_1,\alpha_2))$ implies $\alpha_1 p'<\nu$,
hence $F(x,y,\cdot)\in L^1(B(x,\eta))$ in virtue of the H\"older
inequality and the item (1) of Lemma~\ref{sing}. Similarly
$F(x,y,\cdot)\in L^1(B(y,\eta))$. For the set $Z\equiv
Z(x,y,\eta)=X\setminus (B(x,\eta)\cup B(y,\eta))$ we have
$F(x,y,\cdot)\in L^1(Z(x,y,\eta))$ by the (L2) from the definition
of the classes $\mathcal{K}$ and by H\"older again. Thus,
$F(x,y,\cdot)\in L^1(X)$.

\medskip

(2) Take $r$, $0<r<r_0/2$. Then for $d(x,y)\ge 2r$ we have by
H\"older and (L2):
\begin{multline*}
|J(x,y)|\le\int_{X\setminus B(x,r)}\,|F(x,y,z)|\,dz+
\int_{X\setminus B(y,r)}\,|F(x,y,z)|\,dz\\
\le2\|W\|_p\lfloor K_1\rfloor_{q_1,r}\lfloor K_2\rfloor_{q_2,r}\,.
\end{multline*}
Let now $0<d(x,y)< 2r$. Take a ball $B(a,r)$ with $x,y\in B(a,r)$.
For $p>1$ we have as above, using additionally (L1),
\begin{align*}
|J(x,y)|\le{}& c\,\int_{B(a,2r)}\,d(x,z)^{-\alpha_1}
d(y,z)^{-\alpha_2} |W(z)|\,dz\,\\
&+\int\limits_{X\setminus B(x,r)}\,|F(x,y,z)|\,dz+
\int\limits_{X\setminus B(y,r)}\,|F(x,y,z)|\,dz\\
{}\le{}& c\,\|W\|_p\,\Big(\int\limits_{B(a,2r)}d(x,z)^{-\alpha_1
p\,{}'} d(y,z)^{-\alpha_2 p\,{}'}\,dz\Big)^{1/{p\,{}'}}\\
&+
2\|W\|_p\lfloor K_1\rfloor_{q_1,r}\lfloor K_2\rfloor_{q_2,r}\,,
\end{align*}
with a constant $c>0$. Using now Lemma~\ref{sing}(2), we see that
$J\in \mathcal{K}(\alpha,\infty)$ with required $\alpha$. If
$p=1$, then with necessity $\alpha_1=\alpha_2=1$ and the proof is
obvious.

\medskip

(3) Fix points $x_0,y_0\in X$, $x_0\ne y_0$ and take a number
$\eta$ such that $0<\eta<d(x_0,y_0)/3$. Further fix $\epsilon>0$
and show that $\eta$ can be chosen in such a way that
\begin{itemize}
\item $|J(x,y_0)-J(x_0,y_0)|<\epsilon$ for $x\in B(x_0,\eta/2)$ in
the case (3a);

\item $|J(x_0,y)-J(x_0,y_0)|<\epsilon$ for $y\in B(y_0,\eta/2)$ in
the case (3b);

\item $|J(x,y)-J(x_0,y_0)|<\epsilon$ for $x\in B(x_0,\eta/2)$,
$y\in B(y_0,\eta/2)$ in the case (3c).
\end{itemize}
For this purpose we take a number $R$, $R>2d(x_0,y_0)$, then for
every points $x\in B(x_0,\eta/2)$, $y\in B(y_0,\eta/2)$, the
following estimate takes place
\begin{multline}
|J(x,y)-J(x_0,y_0)|\le\int\limits_{B(x_0,\eta)}|F(x,y,z)|\,dz+
\int\limits_{B(x_0,\eta)}|F(x_0,y_0,z)|\,dz\,+{}\\
\int\limits_{B(y_0,\eta)}|F(x,y,z)|\,dz+
\int\limits_{B(y_0,\eta)}|F(x_0,y_0,z)|\,dz\,+{}\\
 \bigg|\,\,\int\limits_{Z(x_0,y_0,\eta)\cap B(x_0,R)}[F(x,y,z)-F(x_0,y_0,z)]\,dz\,\bigg|+{}\\
 \int\limits_{X\setminus B(x_0,R)}|F(x,y,z)|\,dz\,+\,
\int\limits_{X\setminus B(x_0,R)}| F(x_0,y_0,z)|\,dz\,,
                      \label{aaa}
\end{multline}
where as before $Z(x_0,y_0,\eta)=X\setminus (B(x_0,\eta)\cup
B(y_0,\eta))$. For $z\in B(x_0,\eta)$ we have
$ |F(x,y,z)|\le c\,d(x,z)^{-\alpha_1}\,|W(z)|$,
where $c$ does not depend on $x$, $y$ and $z$. Since
$p\,{}'\alpha_1<\nu$ for $1/p\,{}'+1/p=1$ we have, by
Lemma~\ref{sing}(1) and the H\"older inequality,
\begin{equation}
                     \label{L3.5}
\int_{B(x_0,\eta)}\big|F(x,y,z)\big|\,dz\le
c'\eta^{(\nu/p\,{}')-\alpha_1}\,
\end{equation}
where $c'$ is independent of $x$ and $y$. Similarly,
\begin{equation}
                     \label{L3.5a}
\int_{B(y_0,\eta)}\big|F(x,y,z)\big|\,dz\le
c''\eta^{(\nu/p\,{}')-\alpha_2}\,
\end{equation}
with $c''$ independent of $x$ and $y$ again. We stress that
(\ref{L3.5}) and (\ref{L3.5a}) are valid {\it for all} $x\in
B(x_0,\eta/2)$, $y\in B(y_0,\eta/2)$, in particular, for $x=x_0$,
$y=y_0$. Now we chose $\eta$ such that
$2c'\eta^{(\nu/p\,{}')-\alpha_1}+2c''\eta^{(\nu/p\,{}')-\alpha_2}<\epsilon/3$.
The sum of the last two terms in (\ref{aaa}) are estimated from
above with the help of the H\"older inequality by
\[
2\,\lfloor K_1\rfloor_{q_1,R-d}\lfloor
K_2\rfloor_{q_2,R}\,\|\chi_R W\|_p\,,
\]
where $d=d(x_0,y_0)$ and $\chi_R$ is the characteristic function
of the set $X\setminus B(x_0,R)$. Therefore we can assume by
appropriate choice of $R$ this sum is $<\epsilon/3$. Denoting
$M(\eta,R):=Z(x_0,y_0,\eta)\cap B(x_0,R)$,
it remains to
prove for the obtained $\eta$ and $R$ that the following functions
are continuous:
\begin{itemize}
\item $ B(x_0,\eta/2)\ni x\mapsto
\displaystyle\int_{M(\eta,R)}F(x,y_0,z)\,dz$  in the case (3a);

\item $B(y_0,\eta/2)\ni y\mapsto
\displaystyle\int_{M(\eta,R)}F(x_0,y,z)\,dz$  in the case (3b);

\item $ B(x_0,\eta/2)\times B(y_0,\eta/2)\ni (x,y)\mapsto
\displaystyle\int_{M(\eta,R)}F(x,y,z)\,dz $  in the case~(3c).
\end{itemize}
For this purpose we note that  for $(x,y)\in B(x_0,\eta/2)\times
B(y_0,\eta/2)$ and $z\in Z(x_0,y_0,\eta)\cap B(x_0,R)$ we have the
estimate $\big|F(x,y,z)\big|\le \text{const}\,\big|W(z)\big|$ and
$W\in L^1\big(B(x_0,R)\big)$. Therefore the required continuity
properties follow from the Lebesgue majorization theorem and
conditions (3a)--(3c).

(4) Fix $\eta$ with $0<\eta<r_0$. To prove $F(x,y,\cdot)\in L^1(X)$ we
need to consider only the case $x=y=:x_0$. But in this case we
have with a constant $c>0$ the estimates $|F(x_0,x_0,z)|\le
c\,{d(x_0,z)^{-\alpha_1-\alpha_2}}\,|W(z)|$ for all $z\in
B(x_0,\eta)$, $z\ne x_0$. Therefore, the inclusion
$F(x_0,x_0,\cdot)\in L^1(B(x_0,\eta))$ is a consequence of
Lemma~\ref{sing}(1), the inequality
$(\alpha_1+\alpha_2)q\,{}'<\nu$ and the H\"older inequality. The
inclusion $F(x_0,x_0,\cdot)\in L^1(X\setminus B(x_0,\eta))$
follows from (L2) and the H\"older again.

For proving the properties (4a)--(4c) we proceed as in the proof
of the item (3) and use the notations of this proof. Now we must
consider only the case $x_0=y_0$; in this case we estimate
\begin{multline*}
|J(x,y)-J(x_0,x_0)|\le\int\limits_{B(x_0,\eta)}\big|F(x,y,z)\big|\,dz+
\int\limits_{B(x_0,\eta)}\big|F(x_0,x_0,z)\big|\,dz\,+{}\\
 \bigg|\,\,\int\limits_{B(x_0,R)\setminus B(x_0,\eta)}
 \Big(F(x,y,z)-F(x_0,x_0,z)\Big)\,dz\,
 \bigg|+{}\\
 \int\limits_{X\setminus B(x_0,R)}\big|F(x,y,z)\big|\,dz\,+\,
\int\limits_{X\setminus B(x_0,R)}\big| F(x_0,x_0,z)\big|\,dz\,.
\end{multline*}
The sum of the first two terms has the upper bound of the form
$c'\eta^{(\nu/q\,{}')-\alpha_1-\alpha_2}$ where the exponent is
strictly positive; the sum of the last two terms is estimated by
$
2\,\lfloor K_1\rfloor_{q_1,R-d}\lfloor
K_2\rfloor_{q_2,R}\,\|\chi_R W\|_p\,
$
as before, and to use the Lebesgue majorization theorem again we
have the estimate $\big|F(x,y,z)\big|\le
\text{const}\,\big|W(z)\big|$ for $z\in B(x_0,R)\setminus
B(x_0,\eta)$.
\end{proof}

The suppositions of Lemma~\ref{lemma3} are essential.
Indeed, there holds the following
\begin{prop}\label{prop-kern}
There is a positive symmetric kernel $K\in C^\infty({\mathbb
R}^2)$ such that\\ $(1)$ $K$ is Carleman, moreover, for any
$a\in\RR$ the functions $K(a,\cdot)$ and $K(\cdot,a)$ belong to
$C_0^\infty({\mathbb R})$; $(2)$ $K$ defines a bounded operator
$S$ in $L^2(\mathbb{R})$; $(3)$ for some $f\in L^2({\mathbb R})$
the function $ g(x)=\displaystyle\int_{\mathbb R}K(x,y)f(y)\,dy $
is not equal a.e. to any continuous function on $\mathbb{R}$.
\end{prop}

\begin{proof}
To obtain a kernel $K$ with the required properties we use a
construction from \cite{DSC}. Fix a function $\phi\in
C^\infty(\mathbb{R})$ such that $\phi(x)=0$ if $x\le0$,
$\phi(x)=1$ if $x\ge1$, $\phi\,{}'(x)>0$ if $0<x<1$, and set
$\phi_1(x):=\phi(x+1)\phi(2-x)$, $\phi_2(x)=\phi(x-2)$,
$\psi(x)=\phi(2x)\phi(2-2x)$. Define the kernel
$K(x,y):=M(x,y)+M(y,x)$ with
$M(x,y):=\phi_1(x)\,\phi_2(y)\,L(x,y)$ and
\[
L(x,y):=\begin{cases} 0, & \text{ for } x\le0 \text{ or } y\le0,\\
x^{-1}\psi(y-x^{-1}), & \text{ for } x,y>0,\\
\end{cases}
\]
Let us prove (1). First we
note that $\supp L\subset\{(x,y)\in{\mathbb
R}^2:\,x,y>0,\,xy>1\}$ and that the restriction of $L$ to the set
$\{(x,y)\in{\mathbb R}^2:\,x>0,y>0\}$ is a $C^\infty$-function,
therefore $M\in C^\infty({\mathbb R}^2)$. Denote $U_x=\supp\psi(\cdot-x^{-1})$,
$V_x=\supp\psi(y-{\cdot}^{-1})$. It is easy to see that $U_x\subset
\{y\in{\mathbb R}: \,x,y>0,\,,x^{-1}<y<1+x^{-1}\}$, $V_y\subset\{
x\in{\mathbb R}: \,x,y>0,\,,y^{-1}<x<(y-1)^{-1}\}$, and (1) is
proven.

Let us prove (2); actually we prove that $M$ defines a bounded
operator in $L^2(\mathbb{R})$. Denote $f(y)=\min(1,|y|^{-1})$; due
to the Schur Theorem \cite[Theorem 5.2]{HS} it is sufficient to
prove that
\[
\int_\RR M(x,y)f(y)\,dy\le C_1\,,\quad\quad \int_\RR M(x,y)\,dx\le
C_2\,f(y)
\]
with some constants $C_1,C_2>0$.  We have
\[
\int_{\mathbb R}M(x,y)f(y)\,dy\le
x^{-1}\int_{y\ge2,\,\,y\in U_x}\psi(y-x^{-1})f(y)\,dy\,.
\]
If $y\in U_x$, then $x^{-1}\le y$, therefore $f(y)\le x$. Hence
\[
\int_{\mathbb R}M(x,y)f(y)\,dy\le \int_{\mathbb R
}\psi(y-x^{-1})\,dy=\int_\RR\psi(x)\,dx<\infty\,.
\]
On the other hand, if $y\le2$, then $\displaystyle\int_{\mathbb R}M(x,y)\,dx=0$.
Suppose $y>2$, then
\[
\int_{\mathbb R}M(x,y)\,dx\le \int_{x>0,\,\,x\in
V_y}x^{-1}\psi(y-x^{-1})\,dx\,.
\]
If $x>0$ and $x\in V_y$, then $y^{-1}\le x\le(y-1)^{-1}$, hence
\[
\int_{\mathbb R}M(x,y)\,dx\le
\int_{y^{-1}}^{(y-1)^{-1}}x^{-1}\,dx=\ln(1+(y-1)^{-1})\le
(y-1)^{-1}\le 2y^{-1}\,.
\]
As a result, we have $\displaystyle\int_\RR M(x,y)\,dx\le 2f(y)$
and the item (2) is proven.
To prove (3), we first note that $f\in L^2({\mathbb R})$ and then
show that the function
$g(x)=\displaystyle\int_{\mathbb R}K(x,y)f(y)\,dy$,
(where $f$ is defined above) is piecewise continuous in a
neighborhood of the point $x=0$ and has a jump at this point.
Since $M(\cdot,x)=0$ if $x<2$, it is sufficient to prove that the
function $h(x)=\displaystyle\int_{\mathbb R}M(x,y)f(y)\,dy$,
is piecewise continuous and has a jump at the point $x=0$. It is
clear that $h(x)=0$ if $x<0$ or $x>2$. Let $0<x<1/3$, then
\[
h(x)\ge \int_3^\infty M(x,y)f(y)\,dy=x^{-1}
\int_{y\ge3,\,y\in U_x} \psi(y-x^{-1})f(y)\,dy\,.
\]
If $y>3$ and $y\in U_x$, then $y\le 1+x^{-1}$, therefore, for the
same values of $y$, $f(y)=y^{-1}\ge x(x+1)^{-1}$. Hence, we have
for $0<x<1/3$
\[
h(x)\ge(x+1)^{-1}\int_{y\ge3,\,y\in U_x} \psi(y-x^{-1})\,dy=
(x+1)^{-1}\int_{\mathbb R} \psi(y)\,dy\ge\frac{3}{4}
\int_{\mathbb R} \psi(y)\,dy\,,
\]
and the item (3) is proven.
\end{proof}

Take the kernel $K$ and the function $f$  from
Proposition~\ref{prop-kern}, then setting $K_1=K$, $K_2=1$, $W=f$
in Lemma~\ref{lemma3} we get a discontinuous function $J$, which
demonstrates the importance of the assumptions in
Lemma~\ref{lemma3}.

\section{Norm estimates for the kernels}\label{sec3}
We start with an auxiliary result.
\begin{lem}
                         \label{pos}
Let $V\in\mathcal{P}(X)$ be semi-bounded below: $V\ge -C_V$,
where $C_V\ge0$, then:
\begin{enumerate}
\item[$(1)$] $H_{A}+V$ is semi-bounded below and essentially
self-adjoint on $C_0^\infty(X)$. For every $t>0$ we have
$|e^{-tH_{A\,,V}}| \le
 e^{C_Vt}e^{-tH_0}$ in the sense of point-wise order.
\item[$(2)$] Let $1\le p\le q\le\infty$. Then
$\|e^{-tH_{A,V}}\|_{p,q}\le C_{p,q}\,t^{-\gamma}\,\exp(B_{p,q}t)$,
where  $\displaystyle\gamma=\frac{1}{2}\nu(p^{-1}-q^{-1})$ and
$B_{p,q},\,\,C_{p,q}\ge 0$ (i.e. $H_{A,V}$ obeys the condition
$({\rm S}_{pq})$ from Subsection~\ref{sec-sa} {\rm for all} $p,q$ with $1\le p\le q\le\infty$).

\item[$(3)$] for $t>0$, $e^{-tH_{A,V}}$ is an integral operator
and $|P_{A,V}(x,y;t)|\le e^{C_Vt}\,P_0(x,y;t)$ for a.e. $x,y$.

\item[$(4)$] for $\kappa>0$ and $E<0$ with sufficiently large
$|E|$, the operator $R^\kappa_{A,V}(E)$ has an integral kernel
$G_{A,V}^{(\kappa)}(x,y;E)$ obeying the condition
$|G_{A,V}^{(\kappa)}(x,y;E)|\le G_0^{(\kappa)}(x,y;E+C_V)$. In
particular, at least for $\kappa\ge1$ we have
$G_{A,V}^{(\kappa)}(E) \in\mathcal{K}_0(\alpha,p)$ for all $p$,
$1\le p\le\infty$, where $\alpha=0$ if $\nu=1$, $\alpha$ be an
arbitrary number from the interval $(0,\nu)$ if $\nu=2$ and
$\alpha=\nu-2$ if $\nu\ge3$.
\end{enumerate}
\end{lem}

\begin{proof} (1)It is clear that the operator $H_A+V$ is
semi-bounded below,
therefore it is essentially self-adjoint on $C^\infty_0(X)$ by
Theorem~\ref{thmd}. In particular, $H_{0,V}$ is essentially
self-adjoint on $\mathcal{D}(H_0)\cap \mathcal{D}(V)$. Hence, we
can use the Trotter product formula and for $f\in L^2(X)$ we get
\begin{equation}
                  \label{pos1}
\exp(-tH_{0,V})f=\lim\limits_{n\to
\infty}(\exp(-tH_0/n)\exp(-tV/n))^nf
\end{equation}
with respect to the $L^2$-norm. Eq.~(\ref{pos1}) shows that
$\displaystyle 0\le e^{-tH_{0,V}}f\le e^{C_Vt}e^{-tH_0}f$, if
$f\ge 0$; in virtue of Theorem~\ref{thmf}, the item (1) is proven.

(2) Inequality (\ref{Gri2}) means that
$\sup\{\|e^{-tH_0}\|_{\infty,\infty}:\,t\ge0\}\le 1$. On the other
hand, we obtain from (\ref{Gri1})
\begin{equation}
                      \label{Gri3}
\sup\limits_{x,y} P\,{}_0(x,y\,;t)\le \frac{\tilde
C_P}{\min(t^{\nu/2},1)}
\end{equation}
with $\tilde C_P\ge C_P$. This means that
$\|e^{-tH_0}\|_{1,\infty}\le \tilde C_P\max(t^{-\nu/2},1)$. Using
the Stein interpolation theorem (Theorem~IX.21 from \cite{RS1}) we
finish the proof of the item (2).

(3) Theorem~\ref{thma} and item (2) imply the first statement; the
estimate follows from the estimate in (1).

(4) The existence of integral kernels is a consequence of the item
(2) and Theorem~\ref{thmb}. To get the estimates on the kernels we
use the transformation (\ref{fres}) for the kernels from the item
(3). The last assertion is an immediate consequence of
Lemma~\ref{K0}. \end{proof}

\begin{rem} We stress again that the kernels $P_{A,V}$ and
$G^{(\kappa)}_{A,V}$ are defined not uniquely but only modulo a
negligible function. Moreover, $R^{\kappa}_{A,V}(\zeta)$ can be
not an integral operator for every $\zeta\in\res {H_{A,V}}$; i.e.,
this is the case, if $\kappa=1$, $\nu\ge4$.
\end{rem}

Define kernels $K_\nu(x,y)$,
\begin{equation}
                         \label{Kaker}
K_\nu(x,y)=
\begin{cases}
   d(x,y)^{2-\nu}\,, & \text{if $\nu\ne 2$}\,,\\
    |\log d(x,y)|\,, & \text{if $\nu=2$}\,,\\
\end{cases}
\end{equation}
and for each function $f$ from $L^1_\mathrm{loc}(X)$ and each $r>0$ define
the quantities (\emph{``Kato norms''})
\begin{equation}
                 \label{AA8}
\|f\|^{(r)}_{\rm K}:=\sup\limits_{x\in X}\,\int_{d(x,y)\le
r}\,K_\nu(x,y)\,|f(y)|\,dy\,.
\end{equation}
If $\|f\|^{(r)}_{\rm K}<\infty$ for some $r>0$, then this holds
for any $r>0$.

\begin{lem}
                         \label{Kato}
Let $f\in L^p_{\rm unif}(X)$ where $p=1$ if $\nu=1$ and $p>\nu/2$
otherwise. Then
\[
\lim\limits_{r \downarrow 0}\,\|f\|^{(r)}_{\rm K}\,=0\
\]
uniformly in the unit ball $\|f\|_{p,{\rm unif}}\le1$.
\end{lem}

\begin{proof}
This is an immediate consequence of Lemma~\ref{sing}(1).
\end{proof}

\begin{rem} Lemma~$\ref{Kato}$ means that $L^p_{\rm unif}(X)$
is a subspace of the corresponding ''Kato class'', which can be
defined on the manifold $X$ in the same way as in the case of the
Euclidean space $\mathbb{R}^\nu$~\textup{\cite{CFKS,DvC}}.
\end{rem}

Below we need the following lemmas.

\begin{lem}
                            \label{A1}
Let $F(\rho,t)$ be a measurable function on
$(0,\infty)\times(0,\infty)$ which obeys for each $\rho$ and $t$
the condition
\begin{equation*}
0\le F(\rho\,,t)\le
\frac{1}{\min(t^{\nu/2},1)}\exp\left(-\frac{\rho^2}{a^2t}\right)\,,
\end{equation*}
where $a>0$ is fixed. For $0\le t\le1$ denote
\begin{equation*}
Q(\rho,t)=\int_0^t F(\rho,\,s)\,ds\,.
\end{equation*}
Then with some constant $c_\nu>0$ we have:
\begin{align}
                    \label{LA13}
Q(\rho,\,t)&\le
c_\nu\frac{a^{\nu-2}}{\rho^{\nu-2}}\,\exp\left(-\frac{\rho^2}{2a^2t}\right)\,
& \text{ for } \nu\ge3,\\
                    \label{LA12}
Q(\rho,\,t)&\le
\begin{cases}
\displaystyle
|\log(\rho^2/a^2t)|+1, & \text{ if }\,\, \rho^2<a^2t,\\
\noalign{\medskip} \displaystyle
c_\nu\,\exp\left(-\frac{\rho^2}{2a^2t}\right), & \text{ if }\,\,
\rho^2\ge a^2t\,
\end{cases}& \text{ for } \nu=2,\\
\intertext{and}
                    \label{LA11}
Q(\rho,\,t)&\le
\begin{cases}
\displaystyle
2\sqrt{t}, & \text{ if }\,\, \rho^2<a^2t,\\
\noalign{\medskip} \displaystyle
c_\nu\,\frac{\rho}{a}\exp\left(-\frac{\rho^2}{2a^2t}\right), &
\text{ if }\,\, \rho^2\ge a^2t\,
\end{cases}& \text{ for } \nu=1
\end{align}
\end{lem}

\begin{proof}  By the change of variable we obtain
\begin{multline}
                      \label{A1A}
Q(\rho,\,t)\le \frac{a^{\nu-2}}{\rho^{\nu-2}}
\int\limits_{\rho^2/a^2t}^{+\infty}
\,s^{{\nu/2}-2}\,e^{-s}\,ds\,\\
\le \frac{a^{\nu-2}}{\rho^{\nu-2}}\exp\left(-\frac{\rho^2}{2ta^2}\right)
\int\limits_{\delta^2/a^2t}^{+\infty}
\,s^{{\nu/2}-2}\,e^{-s/2}\,ds\,.
\end{multline}
Denoting
\[
c_\nu:=\int_{0}^{+\infty} \,s^{{\nu/2}-2}\,e^{-s/2}\,ds\,,
\]
we get (\ref{LA13}). Let $\nu\le2$; in the case $\rho^2<a^2t$ we
represent
\begin{align*}
\int_{\rho^2/a^2t}^{+\infty}
\,s^{{\nu/2}-2}\,e^{-s}\,ds&= \int_{\rho^2/a^2t}^{1}
\,s^{{\nu/2}-2}\,e^{-s}\,ds\,+ \int_{1}^{+\infty}
\,s^{{\nu/2}-2}\,e^{-s}\,ds\,\\
&\le \int_{\rho^2/a^2t}^{1} \,s^{{\nu/2}-2}\,ds\,+e^{-1}\,,
\end{align*}
and from the first inequality in (\ref{A1A}) obtain immediately
(\ref{LA12}) and (\ref{LA11}) for the considered case. If
$\rho^2\ge a^2t$ we denote
\[
c_\nu:=\int_{1}^{+\infty} \,s^{{\nu/2}-2}\,e^{-s/2}\,ds\,,
\]
and finish the proof of (\ref{LA12}) and (\ref{LA11}).
\end{proof}

\begin{lem}
                            \label{A2}
Let $W\in L^p_{\rm unif}(X)$ where $p=1$ if $\nu=1$ and $p>\nu/2$
otherwise, and let $P(x,y;t)=F(d(x,y),\,t)$ where $F$ is from
Lemma~\ref{A1}. Then for all sufficiently small $t>0$ there holds
\[
\sup\limits_{x}\,\int_X\,\int_0^t\,P(x,y\,;s)\,
|W(y)|\,ds\,dy<\infty\,.
\]
Moreover,
\[
\lim\limits_{t\downarrow0}\,\sup\limits_{x}\,\int_X\,\int_0^t\,P(x,y\,;s)\,
|W(y)|\,ds\,dy=0
\]
uniformly with respect to $W$ in the unit ball of $L^p_{\rm
unif}(X)$.
\end{lem}


\begin{proof} We can suppose $0<t<1$ and $a \sqrt{t}\le r_0$.
Using the notation of Lemma~\ref{A1} we have
\begin{multline*}
\sup\limits_{x}\,\int_X\,\int_0^t\,P(x,y\,;t)\,
|W(y)|\,dy\,\le \sup\limits_{x}\,\int\limits_{d(x,y)\le a
\sqrt{t}}\,Q(x,y\,;t)\, |W(y)|\,dy\,+{}\\
\sup\limits_{x}\,\int\limits_{a \sqrt{t}<d(x,y)\le a
\sqrt[4]{t}}\,Q(x,y\,;t)\,
|W(y)|\,dy+{}\\\sup\limits_{x}\,\int\limits_{d(x,y)>a
\sqrt[4]{t}}\,Q(x,y\,;t)\, |W(y)|\,dy\,=:F_1(t)+F_2(t)+F_3(t).
\end{multline*}
Consider the function $F_1(t)$. From Lemma~\ref{A1} $F_1(t)\le
2\sqrt{t}\|W\|_{1,{\rm unif}}$ if $\nu=1$. For $\nu\ge3$ we obtain
$F_1(t)\le {\rm const}\,\|W\|^{(r)}_{\rm K}$ where $r=a\sqrt{t}$.
Since $d(x,y)\le a\sqrt{t}$ implies $|\log a^2t|\le|\log
d(x,y)^2|$, in the case $\nu=2$ the inequality (\ref{LA12})
implies $F_1(t)\le {\rm const}\,\|W\|^{(r)}_{\rm K}$ from
Lemma~\ref{A1} with $r=a\sqrt{t}$ again. Hence, $F_1(t)\to 0$ as
$t\to0$ uniformly in the unit ball of $L^p_{\rm unif}(X)$ due to
Lemma~\ref{Kato}.

In the region $d(x,y)>a\sqrt{t}$ we have according to
Lemma~\ref{A1} (in the case $\nu=2$ we consider sufficiently small
$t$) $Q(x,y\,;t)\le {\rm const} K_\nu(x,y)$
with kernels $K_\nu$ from (\ref{Kaker}). Hence, $F_2(t)\le {\rm
const} \|W\|_{\rm K}^{(r)}$ with $r=a\sqrt[4]{t}$, and by
Lemma~\ref{Kato} $F_2(t)\to0$ uniformly in the unit ball of
$L^p_{\rm unif}(X)$ as $t\to0$.

Finally, consider $F_3(t)$. Choose now $t_0$, $t_0>0$, such that
$(5t_0a^2)^{-1}>\theta_X$. According to Lemma~\ref{A1} we have for
$t<t_0$ in the region $d(x,y)>a\sqrt[4]{t}$:
\begin{equation}
          \label{AA12a}
\begin{aligned}
Q(x,y\,;t)&\le {\rm const}\,
d(x,y)^{2-\nu}\,\exp\left(-\frac{d(x,y)^2}{4t_0a^2}\right)
\exp\left(-\frac{1}{4\sqrt{t}}\right)\,\\
&\le{\rm const}\,
d(x,y)^{1-\nu}\,\exp\left(-\frac{d(x,y)^2}{5t_0a^2}\right)
\exp\left(-\frac{1}{4\sqrt{t}}\right)\,\\
&\le {\rm const}\,
t^{(1-\nu)/4}\,\exp\left(-\frac{d(x,y)^2}{5t_0a^2}\right)
\exp\left(-\frac{1}{4\sqrt{t}}\right)\,.
\end{aligned}
\end{equation}

In virtue of Lemma~\ref{exp}, for each $x\in X$ the function
\[
g_x(y)=\exp\left(-\frac{d(x,y)^2}{5\sqrt{t_0}a^2}\right)|W(y)|
\]
belongs to $L^1(X)$ and $\|g_x\|_1\le c' \|W\|_{1,\,\,\rm unif}$
where $c'$ is independent of $t$ and $x$. Therefore, we have from
(\ref{AA12a})
\[
F_3(t)\le {\rm const}\,t^{(1-\nu)/4}
\exp\left(-\frac{1}{4\sqrt{t}}\right)\|W\|_{1,{\rm unif}}\,,
\]
thus $F_3(t)\to0$ as $t\to0$ uniformly in the unit ball of
$L^1_{\rm unif}$ and hence, of $L^p_{\rm unif}$.
\end{proof}

The following theorem is the main result of the section.
\begin{thm}
                            \label{Th1}
Let $U\in\mathcal{P}(X)$, then the following assertions are true
\begin{enumerate}
\item[$(1)$] $H_A+U$ is essentially self-adjoint and semi-bounded
below on $C_0^\infty(X)$.

\item[$(2)$] Let $1\le p\le q\le\infty$. Then
$\|e^{-tH_{A,V}}\|_{p,q}\le C_{p,q}\,t^{-\gamma}\,\exp(B_{p,q}t)$,
where  $\displaystyle\gamma=\frac{1}{2}\nu(p^{-1}-q^{-1})$ and
$B_{p,q},\,\,C_{p,q}\ge 0$ (i.e. $H_{A,V}$ obeys the condition
$({\rm S}_{pq})$ from Subsection~\ref{sec-sa} {\rm for all} $p,q$ with $1\le p\le q\le\infty$).
\item[$(3)$] There are $C>0$ and $a>0$ such that for any compact
sets $K_1,K_2\subset X$ with $d:={\rm dist}\,(K_1,K_2)>0$ we have
for all $t$, $0<t<1$,
\[
\|\chi_1\,e^{-tH_{A,U}}\,\chi_2\|_{1,\infty}\,\le
Ct^{-\nu/2}e^{-d^2/a^2t}\,,
\]
where $\chi_j$ is the characteristic function of $K_j$, $j=1,2$.
\item[$(4)$] For $\zeta\in\res{H_{A,U}}$ with $\Re\zeta<0$ and
sufficiently large $|\Re\zeta|$, the kernel
$G_{A,U}^{(\kappa)}(\zeta):=G_{A,U}^{(\kappa)}(\cdot,\cdot\,;\zeta)$
of $R_{A,U}^\kappa(\zeta)$ exists for each $\kappa>0$ and
$G_{A,U}^{(\kappa)}(\zeta)\in \mathcal{K}(\alpha,q)$ where $q$,
$1\le q\le\infty$, is arbitrary, and $\alpha=\nu-2\kappa$ for
$\kappa<\nu/2$, $0<\alpha<\nu$ is arbitrary for $\kappa=\nu/2$,
and $\alpha=0$ for $\kappa>\nu/2$.
\end{enumerate}
\end{thm}

\begin{proof} We can represent $U$ in the form $U=V-W$,
where $V\in \mathcal{P}(X)$ and is semi-bounded below, and
$W=\sum_{j=1}^n\,W_j$ where $W_j\ge0$, $W_j\in L^{p_j}(X)$ with
$2\le p_j<\infty$ if $\nu\le3$ and $\nu/2<p_j<\infty$ otherwise.

\medskip

(1) Since
\begin{multline*}
\|R^{1/2}_{A,V}(E)WR^{1/2}_{A,V}\|_{2,2}\,\le\,
\sum_{j=1}^n\|R^{1/2}_{A,V}(E)W_jR^{1/2}_{A,V}\|_{2,2}\,\\
=
\sum_{j=1}^n\|\,W_j^{1/2}R^{1/2}_{A,V}(E)\|^2_{2,2}=
\sum_{j=1}^n\|\,W_j^{1/2}R_{A,V}(E)W_j^{1/2}\|_{2,2}\,,
\end{multline*}
we have according to Corollary~\ref{corol4}(2b) that
$\|R^{1/2}_{A,V}(E)WR^{1/2}_{A,V}(E)\|_{2,2}\to0$ as
$E\to-\infty$. Therefore, $W$ is form-bounded with respect to
$H_{A,V}$ and the item (1) follows from Theorem~\ref{thmd}.

\medskip

(2) As shown in the proof of the inequality (B11) in
\cite{Sim}, it is sufficient to prove the following relations:
\begin{description}
\item[(R1)] there is $T>0$ such that
$\sup_{0\le t\le
T}\,\|e^{-tH_{A,U}}\|_{\infty,\infty}<\infty$\,;
\item[(R2)] there are $\tilde B>0$ and $\tilde C>0$ such that
$\|e^{-tH_{A,U}}\|_{2,\infty}\le \tilde C\,t^{\nu/4}\,\exp(\tilde
B\,t)$ for all $t>0$.
\end{description}
Taking into account Theorem~\ref{thmf}, we have to prove (R1) and
(R2) for the case $A=0$ only. For this purpose we use the ideas of
the proofs of Theorem~B.1.1 from \cite{Sim} and Theorem~2.1 from
\cite{CFKS}. Let us start with (R1). First of all, from
Lemmas~\ref{pos} and \ref{A2} we see that
\begin{equation*}
\lim\limits_{t\downarrow
0}\Big\|\int_0^t\,e^{-sH_{0,V}}W\,ds\,\Big\|_{\infty,\infty}=0\,
\end{equation*}
uniformly in $W$ from the unit ball of $L^p(X)$. Let
$\displaystyle W^{(n)}(x)=\sum_{j=1}^n\,\min(W_j(x),n)$ and
$H_n=H_{0,V}-W^{(n)}$. Since $0\le W^{(n)}\le W$ for all $n$, we
can find constants $T>0$ and $\eta$, $0<\eta<1$, such that
\[
\Big\|\int_0^T\,e^{-sH_{0,V}}W^{(n)}\,ds\,\Big\|_{\infty,\infty}\le\eta
\]
for all $n$. Fix now $t$, $0<t<T$. Using the Dyson--Phillips
expansion we show that $\|e^{-tH_n}\|_{\infty,\infty} \le
(1-\eta)^{-1}$, see the proof of Theorem~2.1 from \cite{CFKS}. On
the other hand, $H_n:=H_{0,V}-W^{(n)}$ tends to $H_{0,U}$ in the
strong resolvent sense \cite[Theorem~VIII.25]{RS1}. Let $\phi\in
L^2(X)$, $\|\phi\|_{\infty}\le1$, then
$\|e^{-tH_n}\phi-e^{-tH_{0,U}}\phi\|_2\to 0$ and
$\|e^{-tH_n}\phi\|_\infty\le (1-\eta)^{-1}$ for all $n$. We can
extract a subsequence $(e^{-tH_{n_k}}\phi)_{k\ge1}$ which tends to
$e^{-tH_{0,U}}\phi$ a.e., hence
$\|e^{-tH_{0,U}}\phi\|_\infty\le(1-\eta)^{-1}$ and the statement
(R1) is proven.

To proceed further we need the following ``Schwarz inequality''
\begin{equation}
                            \label{sg2}
\big|(e^{-tH_{0,U}}f)(x)\big|^2\le
\big|e^{-t(H_{0,V}-2W)}1(x)\big|\,\big|e^{-tH_{0,V}}|f|^2(x)\big|\,\quad
\text{for a.e. }x\in X,
\end{equation}
where $f\in L^2(X)$. Replaced $W$ by cut-off functions $W^{(n)}$
defined above we can repeat the proof of Lemma~6.4 from
\cite{DHSV} (see also proof of Theorem~2.1 from \cite{CFKS}) to
derive (\ref{sg2}) with $W^{(n)}$ instead of $W$, and then extend
this inequality to $W$ by the limiting considerations above. The
property (R1) implies that for all $t>0$  we have
$\|e^{-t(H_{0,V}-2W)}\|_{\infty,\infty}\le C_1e^{tB_1}$ with some
$B_1,\,C_1>0$ (see the mentioned prof from \cite{CFKS}), whereas
Lemma~\ref{pos}(2) and inequality (\ref{Gri3}) imply for $f\in
L^2(X)$, $\|f\|\le1$,
\[
\big\|\,e^{-tH_{0,V}}|f|^2\big\|_{\infty}\le\frac{C_2}{\min(t^{\nu/2},1)}e^{B_2t}\,,
\]
with some $B_2,\,C_2>0$. Using (\ref{sg2}) we finish checking the
property (R2) and, therefore, the proof of the item (2).

\medskip

(3) To prove this item it is sufficient to follow the
proof of Proposition~B.4.2 from \cite{Sim}.

\medskip

(4) The existence of the integral kernels follows from
Theorem~\ref{thmb}(3). Arguing further as in the proof of
Lemma~B.7.6 in~\cite{Sim}, we can show that for each $d>0$ there
is a constant $c_d>0$ such that for $\zeta\in\res{H_{A,U}}$, where
$\Re\zeta<0$ and $|\Re\zeta|$ is sufficiently large we have
$|G_{A,U}(x,y;\zeta)|\le c_d$ for $d(x,y)\ge d$. Moreover, if
$\nu\ne2\kappa$, then $|G_{A,U}(x,y;\zeta)|\le c_d\,d(x,y)^\alpha$
for $d(x,y)\le d$ with $\alpha$ given in the item~(4) of the
theorem. In the case $\kappa=\nu/2$ it is sufficient to replace
the inequality in the item (3) by
$\|\chi_1\,e^{-tH_{A,U}}\,\chi_2\|_{1,\infty}\,\le
Ct^{-(\nu+\epsilon)/2}e^{-d^2/b^2t}$ with $\epsilon>0$, $b>a$, and
repeat the arguments of the proof of Theorem~B.4.3 from
\cite{Sim}. Thus, we show that
$G_{A,U}(\zeta)\in\mathcal{K}(\alpha,\infty)$ for noted $\zeta$
and $\alpha$.

According to Theorems~\ref{thma} and \ref{thmb}, we have $\lfloor
G_{A,U}^{(\kappa)}\rfloor_{p\,{}',r}<\infty$ for every $r<r_0$ if
$p^{-1}<2\kappa/\nu$ if $p^{-1}<2\kappa/\nu$. This condition is
satisfies, if $p=\infty$, hence, if $p\,{}'=1$. Therefore,
$G_{A,U}(\zeta)\in\mathcal{K}(\alpha,1)$ for $\zeta$ and $\alpha$
as above. Thus, by the property (K3) of the classes $\mathcal{K}$
(see Section~\ref{cond-pot}) the theorem is proved.
\end{proof}

\section{Continuity of the kernels}\label{sec4}

Before stating the main result of this section we prove the
following lemma.

\begin{lem}
                           \label{lmain}
Let $f$ be a real-valued function from $L^p_{\rm loc}(X)$, where
$1\le p<\infty$, and $f\ge c$ with a constant $c\in \mathbb{R}$.
Then there exists a real-valued function $g$ from $C^\infty(X)$
such that $g\ge c$ and $\displaystyle f-g\in L^q(X)$ for all $1\le
q\le p$.
\end{lem}

\begin{proof}
Fix $a\in X$ and for integers $n$, $n\ge1$, denote
$Y_n=B(a,n)\setminus\overline{B(a,n-1)}$. Fix a real sequence
$\alpha_n$, $\alpha_n>0$, such that $\sum \alpha_n\le1$ and denote
by $f_n$ the restriction of $f$ to the set $Y_n$. Since the
measure of $Y_n$ is finite, for every $n$ we can find a
real-valued function $g_n$, $g_n\in C_0^\infty(X)$, such that
$g_n\ge c$, ${\rm supp}\,(g_n)\subset Y_n$, and
$\max(\|f_n-g_n\|^p_p,\,\|f_n-g_n\|_1)\,\le \alpha_n$. Since the
family $(Y_n)$ is locally finite, the point-wise sum $g=\sum g_n$
exists and $g\in C^\infty(X)$. It is clear that $g\ge c$ and
$\max(\|f-g\|_p\,,\|f-g\|_1)\,\le1$, i.e., $f-g\in L^p(X)\cap
L^1(X)$; hence, $\displaystyle f-g\in L^q(X)$ for all $1\le q\le
p$\,.
\end{proof}

Now we are in position to prove the main result of the paper.

\begin{thm}
                          \label{main}
Let a potential $U$, $U\in \mathcal{P}(X)$, be given.
\begin{enumerate}

\item[$(1)$] For $t>0$ the operator $e^{-tH_{A,U}}$ has an
integral kernel $P_{A,U}(x,y\,;\,t)$ which is jointly continuous
in $X\times X\times(0,\infty)$.

\item[$(2)$] For any {\rm bounded} Borel set $S\subset
\mathbb{R}$, the corresponding spectral projection for $H_{A,U}$
has a continuous in $X\times X$ integral kernel.

\item[$(3)$] Let $\kappa>0$ and $\zeta\in \res{H_{A,U}}$. Then the
Green function $G^{(\kappa)}_{A,U}(\cdot,\cdot;\zeta)$ is
continuous in $(X\times X)\setminus D$ if one of the following
conditions is valid:
\begin{itemize}
\item[\rm(a)] $\Re\zeta<0$ and $|\Re\zeta|$ is
sufficiently large,
\item[\rm(b)] $\kappa$ is an integer.
\end{itemize}
Moreover, if
$\kappa>\nu/4$, then under these conditions
$G_{A,U}^{(\kappa)}(\cdot,\cdot;\zeta)$ is a Carleman kernel with
the properties $\rm(C1)$ and $\rm(C2)$ from Subsection~\ref{sec-sa}; in
particular, the image of $R^\kappa(\zeta)$ consists of continuous
functions.

\item[$(4)$] If $f$ is a Borel function on $\spec{H_{A,U}}$
obeying the condition $|f(\xi)|\le b(|\xi|+1)^{-\kappa}$ with some
$b>0$ and $\kappa>\nu/2$, then the operator $f(H_{A,U})$ has an
integral kernel $F(x,y)$ which is continuous on $X\times X$.
Moreover $\sup\,\{|F(x,y)|:\,x,y\in X\}\le C\,b<\infty$ where $C$
depends only on~$\kappa$.

\item[$(5)$] If $\kappa>\nu/2$, then for all
$\zeta\in\res{H_{A,U}}$ the kernel
$G^{(\kappa)}_{A,U}(\cdot,\cdot;\zeta)$ is a bounded continuous
function on the whole space $X\times X$.


\item[$(6)$] Each eigenfunction of $H_{A,U}$ is bounded and
continuous.

\item[$(7)$] Let $k$ be an integer, $k\ge1$. Then the map
$\zeta\mapsto G^{(k)}_{A,U}(x,y\,;\,\zeta)$ is holomorphic in
$\res{H_{A,U}}$ for all $x,y\in X$ if $k>\nu/2$, and for $x\ne y$
otherwise. Moreover $\partial
G^{(k)}_{A,U}(x,y\,;\,\zeta)/\partial\zeta=kG^{(k+1)}_{A,U}(x,y\,;\,\zeta)$
for $(x,y)$ above.
\end{enumerate}
\end{thm}

\begin{proof} Using Lemma~\ref{lmain} we represent
$U$ in the form $U=V+W$, where $V$ and $W$ have the properties

\begin{gather}
      \label{MT1}
V\in C^\infty(X) \text{ and is semi-bounded below;}\\
    \label{MT2}
    \begin{gathered}
W=\sum\nolimits_{j=0}^n\,W_j, \quad W_j\in L^{p_j}(X),\\
2\le p_j<\infty \text{ if } \nu\le3 \text{ and } \nu/2<p_j<\infty
\text{ otherwise, } 0\le j\le n.
\end{gathered}
\end{gather}
Let $\kappa$ be any strictly positive number; denote by
$\alpha_\kappa$ the number $\nu-2\kappa$ if $\kappa<\nu/2$, an
arbitrary number from the interval $(0,\nu)$ if $\kappa=\nu/2$,
and $0$ if $\kappa>\nu/2$. Then we have by Theorem~\ref{Th1}(4)
and the properties (K2) and (K3) from Section~\ref{cond-pot}
\begin{equation}
     \label{MT3}
     \parbox{105mm}{for $E<0$ with sufficiently large $|E|$ the kernels
$G^{(\kappa)}_{A,V}(\cdot,\cdot;E)$ and
$G^{(\kappa)}_{A,U}(\cdot,\cdot;E)$ exist and belong to all the
classes $\mathcal{K}(\beta,q)$ with $1\le q\le\infty$,
$\alpha_\kappa\le\beta<\nu$\,.}
\end{equation}
Moreover, by Lemma~\ref{pos}(4),
\begin{equation}
       \label{MT4}
       \parbox{105mm}{for $E<0$ with sufficiently large $|E|$ and for
$\kappa\ge1$ we have
$G^{(\kappa)}_{A,V}(\cdot,\cdot;E)\in\mathcal{K}_0(\beta,q)$ for
every $q$, $1\le q<\infty$ and $\beta$,
$\alpha_\kappa\le\beta<\nu$\,.}
\end{equation}
Further, by virtue of \eqref{MT1}, we have the following continuity
properties:
\begin{equation}
      \label{MT5}
      \parbox{105mm}{$G_{A,V}(x,y;\zeta)$ can be chosen from
$C^\infty\big((X\times X)\setminus D\big)$ if $\nu\ge 2$ and from
$C^\infty(X\times X)$ if $\nu=1$.}
\end{equation}
The first statement in \eqref{MT5} follows from the standard elliptic
regularity considerations \cite{Shu}; the second one can be found
in \cite{Nai}.

Now we show that for $E<0$ with sufficiently large $|E|$ and for
every integer $k$, $k\ge1$ there holds
\begin{gather}
                      \label{Resk}
R^k_{A,U}(E)=R^{k-1}_{A,U}(E)R_{A,V}(E)-R^k_{A,U}(E)WR_{A,V}(E)\,,\\
                      \label{Resik}
R^k_{A,U}(E)=R_{A,V}(E)R^{k-1}_{A,U}(E)-R_{A,V}(E)WR^k_{A,U}(E)\,.
\end{gather}
Passing on to adjoint operators we derive \eqref{Resik} from
\eqref{Resk}, therefore, we consider \eqref{Resk} only. Obviously,
it is sufficient to prove \eqref{Resk} for the case $k=1$. Using
item (2b) from Corollary~\ref{corol4} and
Theorem~\ref{Th1} we get $\|\,|W|^{1/2}R^{1/2}_{A,V}(E)\|_{2,2}=
\|R^{1/2}_{A,V}(E)\,|W|\,R^{1/2}_{A,V}(E)\|_{2,2}^{1/2}<\infty$
and similarly $\|R^{1/2}_{A,U}(E)\,|W|^{1/2}\,\|_{2,2}<\infty$.
Denote, as usual,
\[
\sign W(x)=
\begin{cases} \displaystyle\frac{W(x)}{|W(x)|}\,,
     & \text{ if }\, W(x)\ne0\,,\\
\noalign{\medskip}
     0,\, & \text{ otherwise}\,.
\end{cases}
\]
Then $\|\sign W\,|W|^{1/2}R^{1/2}_{A,V}(E)\|_{2,2}<\infty$,
therefore $\|R^{1/2}_{A,U}(E)WR^{1/2}_{A,V}(E)\|_{2,2}<\infty$,
hence $\|R_{A,U}(E)WR_{A,V}(E)\|_{2,2}<\infty$. It remains to
prove that both of the sides of the equation
\begin{equation}
                      \label{Res}
R_{A,U}(E)=R_{A,V}(E)-R_{A,U}(E)WR_{A,V}(E)\,.
\end{equation}
coincide on a dense subset in $L^2(X)$. Consider functions
$f=(H_{A,V}-E)\phi$ where $\phi$ runs over $C^\infty_0(X)$; these
functions form a dense subset since $H_{A,V}$ is essentially
self-adjoint on $C^\infty_0(X)$. Further, $\phi\in
\mathcal{D}(H_{A,U})$ and $W\phi\in L^2(X)$, therefore $R_{A,U}f=
R_{A,U}((H_{A,U}-E)\phi-W\phi)=\phi-R_{A,U}W\phi$. Since
$\phi=R_{A,V}\,f$, we get the result. Worth noting that
(\ref{Res}) is nothing else than the Lippmann--Schwinger equation
for the potential $W$.

Using \eqref{MT2} and \eqref{MT3} we get with the help of Lemma~\ref{lemma3}
that for $x\ne y$ and $k\ge1$
\begin{equation}
                         \label{ain1}
G^{(k)}_{A,U}(x,\cdot;E)W(\cdot)\,G_{A,V}(\cdot,y;E)\in L^1(X\times
X)\,,
\end{equation}
\[
G_{A,V}(x,\cdot;E)W(\cdot)\,G^{(k)}_{A,U}(\cdot,y;E)\in L^1(X\times
X)\,
\]
for all $E<0$ with sufficiently large $|E|$. Similarly, using
Lemma~\ref{lemma3} with $W\equiv 1$, we get for $k\ge2$ and, for
the same $E$,
\[
G^{(k-1)}_{A,U}(x,\cdot;E)G_{A,V}(\cdot,y;E),\quad
G_{A,V}(x,\cdot;E)G^{(k-1)}_{A,U}(\cdot,y;E)\in L^1(X\times X)\,.
\]
Therefore, the following functions are well defined:
\[
J^{(k)}_1(x,y;E):=\int_X\,G^{(k)}_{A,U}(x,z;E)W(z)G_{A,V}(z,y;E)\,dz\,,
\]
\begin{equation}
                          \label{wd1}
J^{(k)}_2(x,y;E):=\int_X\,G_{A,V}(x,z;E)W(z)G^{(k)}_{A,U}(z,y;E)\,dz\,,
\end{equation}
for $k\ge1$, and
\begin{gather}
L^{(k)}_1(x,y;E):=\int_X\,G^{(k-1)}_{A,U}(x,z;E)G_{A,V}(z,y;E)\,dz\,,\notag\\
                          \label{wd2}
L^{(k)}_2(x,y;E):=\int_X\,G_{A,V}(x,z;E)G^{(k-1)}_{A,U}(z,y;E)\,dz\,,
\end{gather}
for $k\ge2$. Moreover, the integrals in (\ref{wd1}) and
(\ref{wd2}) converge absolutely. Denote
$L^{(1)}_j(x,y;E):=G_{A,V}(x,y;E)$ for $j=1,2$. We show that for
all $k\ge1$ the functions $L^{(k)}_1(E)-J^{(k)}_1(E)$ and
$L^{(k)}_2(E)-J^{(k)}_2(E)$ are the integral kernels of
$R^{(k)}_{A,U}(E)$, i.e.
\begin{equation}
                      \label{GK}
G^{(k)}_{A,U}(x,y;E)=L^{(k)}_j(x,y;E)-J^{(k)}_j(x,y;E)\,,
\end{equation}
for a.e. $(x,y)\in L^1(X\times X)$ $(j=1,2)$. By the item (2) of
Lemma~\ref{lemma3} and the property (K5) from
Section~\ref{cond-pot} all the kernels $J^{(k)}_j$ and $L^{(k)}_j$
belong to $L^1_{\rm loc}(X\times X)$. According to (\ref{Resk})
and (\ref{Resik}) it remains to show that for $\phi,\psi\in
C_0^\infty(X)$, $k\ge1$ and $j=1,2$
\begin{equation*}
\langle\psi\,|\,R^k_{A,U}(E)\phi\rangle=
\int_{X\times
X}\,L^{(k)}_j(x,y)\overline{\psi(x)}\phi(y)\,dxdy-
\int_X\,J^{(k)}_j(x,y)\overline{\psi(x)}\phi(y)\,dxdy\,.
\end{equation*}
Firstly we show that
\begin{equation}
                        \label{ord}
\big\langle\psi\,\big|\,R^k_{A,U}(E)WR_{A,V}(E)\phi\big\rangle=
\int_X\,J^{(k)}_1(x,y)\overline{\psi(x)}\phi(y)\,dxdy\,.
\end{equation}
It was shown by the proof of Equation~(\ref{Res}) that
$\|\,|W|^{1/2}R^{1/2}_{A,V}(E)\|_{2,2}<\infty$ and
$\|\,|W|^{1/2}R^{1/2}_{A,U}(E)\|_{2,2}<\infty$, therefore
$\|\,\sign W\,|W|^{1/2}R_{A,V}(E)\|_{2,2}<\infty$ and
$\|\,|W|^{1/2}R^{k}_{A,U}(E)\|_{2,2}<\infty$. This means that the
functions
\[
f_1(z):=\sign W(z)\,|W|^{1/2}(z)\,\int_X\,G_{A,V}(z,x;E)\,\phi(x)\,dx\,
\]
and
\[
f_2(z):=|W|^{1/2}(z)\,\int_X\,G^{(k)}_{A,U}(z,y;E)\,\psi(y)\,dy\,
\]
are from $L^2(X)$. It is clear that
$\langle\psi\,|\,R^k_{A,U}(E)WR_{A,V}(E)\phi\rangle=\langle
f_2\,|\,f_1\rangle$. In virtue of (\ref{ain1}) and absolute
convergence of integrals (\ref{wd1}), we can change order of
integration in the integral expression for $\langle
f_2\,|\,f_1\rangle$ and obtain (\ref{ord}). Similarly we prove
that
\[
\big\langle\psi\,\big|\,R^{k-1}_{A,U}(E)R_{A,V}(E)\phi\big\rangle=
\int_X\,L^{(k)}_1(x,y)\overline{\psi(x)}\phi(y)\,dxdy\,.
\]
Hence, \eqref{GK} is proved for $j=1$. The case $j=2$ is reduced
to $j=1$ by the simple consideration:
\[
\big\langle\psi\,\big|\,R^k_{A,U}(E)\phi\big\rangle= \big\langle
R^{k-1}_{A,U}(E)R_{A,V}(E)\psi\,\big|\,\phi\big\rangle-\big\langle
R^k_{A,U}(E)WR_{A,V}(E)\psi\,\big|\,\phi\big\rangle.
\]
{}From \eqref{MT2} and \eqref{MT3} it is easy to see that conditions (a) and (b)
of Lemma~\ref{lemma3} are satisfied with $K_1=G_{A,U}^{(k)}(E)$
$(k\ge1)$ and $K_2=G_{A,V}(E)$ with $W$ given in \eqref{MT2}. The same
is true for $W\equiv 1$ in Lemma~\ref{lemma3}. Moreover, in virtue
of \eqref{MT4} the functions $K_1$, $K_2$, and $W$ satisfy the
additional conditions from the item (3) of the lemma and if
$k>\nu/2$ they satisfy the additional conditions of the item (4).
Therefore, by \eqref{MT5} and Lemma~\ref{lemma3}(3b) (or
item (4b), if $k>\nu/2$), for all $x\in X$, the function
$G_{A,U}^{(k)}(x,\cdot;E)$ is continuous in $X\setminus\{x\}$
(respectively, in $X$) for all $k\ge1$. Now taking
$K_1=G_{A,V}(E)$, $K_2=G_{A,U}^{(k)}(E)$ and using the item (3c)
of Lemma~\ref{lemma3}, we get the first statement of the following
assertion concerning the properties of the kernels
$G_{A,U}^{(k)}(E)$:

\begin{equation}
     \label{MT6}
          \parbox{105mm}{$G_{A,U}^{(k)}(E)$ is continuous in $(X\times
X)\setminus D$ for all $k\ge1$, and in $X\times X$ for all
$k>\nu/2$. Moreover, if $k>\nu/2$, then $G_{A,U}^{(k)}(E)$ is a
Carleman kernel obeying the conditions (C1) and (C2) from
Subsection~\ref{sec-sa}}
\end{equation}
To prove the second statement we note firstly that
$G_{A,U}^{(k)}(E)$ is a Carleman kernel for $k>\nu/2$ by
Theorem~\ref{thmb}. Further,
\[
\int_X\,|G_{A,U}^{(k)}(x,y\,;E)|^2\,dy=G_{A,U}^{(2k)}(x,x\,;E)
\]
and $G_{A,U}^{(2k)}(x,x\,;E)$ is continuous in $x$. Hence, the
property (C2) is valid. The property (C1) we get from
Lemma~\ref{lemma3} if we set $K_1=G_{A,U}^{(k)}(E)$, $W=f$,
$K_2\equiv 1$ and take into consideration \eqref{MT4} and \eqref{MT6}.

Since $\|R^{\kappa}_{A,U}(\zeta)\|_{2,\infty}<\infty$ by
Theorems~\ref{thmb}(1) and~\ref{Th1} we can apply the item~(2) of
Theorem~\ref{thmc} and obtain the following assertion:

\begin{equation}
\label{MT7}
\parbox{105mm}{There is an integer $n>1$ such that for any
Borel function $f$ defined on the spectrum of $H_{A,U}$ and having
the property $\big|f(\xi)\big|\le b\big(|\xi|+1\big)^{-n}$ with $b>0$, the
operator $f(H_{A,U})$ has a Carleman continuous kernel $F(x,y)$
with property $\big|F(x,y)\big|\le bc^2$, where $c$ is independent of
$f$.}
\end{equation}
The assertion \eqref{MT7} allows us to deduce all the items
(1)--(7) of the theorem step by step.

(1) Consider for $t>0$ the function $f(\xi)=e^{-t\xi}$.
It is clear that $|f(\xi)|\le b(1+|\xi|)^{-n}$ with some $b>0$,
therefore by \eqref{MT7}, the operator $\exp(-tH_{A,U})$ has a kernel
$P_{A,U}(x,y\,;t)$ which is jointly continuous in $x,y\in X$ at
any fixed $t$, $t>0$. Fix now any $t_0>0$, then in a neighborhood
of $t_0$ we have the estimate $|e^{-t\xi}-e^{-t_0\xi}|\le b(t)
(1+|\xi|)^{-n}$, where $b(t)\to 0$ if $t\to t_0$. By~\eqref{MT7},
$|P_{A,U}(x,y\,;t)-P_{A,U}(x,y\,;t_0)|\le cb(t)$. Now using
the continuity of $P_{A,U}(x,y\,;t)$ with respect to $(x,y)$ we
complete the proof of (1).

(2) This is an immediate consequence of \eqref{MT7}.

(3) According to the equation (\ref{fres}) we have for
$x\ne y$ that
\begin{equation}
                      \label{fresg}
G_{A,U}^{(\kappa)}(x,y;\zeta)=
\frac{1}{\Gamma(\kappa)}\int_0^{\infty}\,P_{A,U}(x,y\,;t)e^{t\zeta}\,t^{\kappa-1}\,dt,
\end{equation}
if $\Re\,\zeta<0$ with sufficiently large $|\Re\,\zeta|$ and if
the integral in (\ref{fresg}) converges absolutely and locally
uniformly in $(X\times X)\setminus D$. Using the item (2) of
Theorem~\ref{Th1} with $p=1$ and $q=\infty$ for $t\ge1$ and the
item (3) for $0<t<1$ we see that the integral (\ref{fresg})
converges absolutely and locally uniformly in $(X\times
X)\setminus D$ if $\Re\,\zeta<B_{1,\infty}$. Therefore, in the
case (a) the proof is completed.

To prove the item in the case (b), we take $n>\nu/2$ from \eqref{MT7}
and fix $E_0<0$ such that the kernels $G^{(1)}_{A,U}(E_0)$, \dots,
$G^{(n)}_{A,U}(E_0)$ are continuous in $X\times X\setminus D$ and
consider an arbitrary $\zeta\in\res{H_{A,U}}$. Let
$f_\zeta(\xi)=(\xi-E_0)^{-n}(\xi-\zeta)^{-1}$ for
$\xi\in\spec{H_{A,U}}$. Then $|f_\zeta(\xi)|\le
b(\zeta)(|\xi|+1)^{-n}$ for all $\xi$ where $b(\zeta)>0$ is
locally bounded in $\zeta$ from $\res{H_{A,U}}$. Using the
identity
\begin{equation}
                          \label{Res.1}
\frac{1}{\xi-\zeta}=\frac{1}{\xi-E_0}+\frac{\zeta-E_0}{(\xi-E_0)^2}+\ldots+
\frac{(\zeta-E_0)^{n-1}}{(\xi-E_0)^{n}}+
\frac{(\zeta-E_0)^{n}}{(\xi-E_0)^{n}(\xi-\zeta)}\,,
\end{equation}
we obtain the part (b) for $\kappa=1$ from the part (a) and the
assertion \eqref{MT7}. To get these items for any positive integer
$\kappa$ it is sufficient to consider the $\kappa$-th power of
both the sides of (\ref{Res.1}), and represent the right-hand side
of the obtained expression as the sum of products of terms in the
right-hand side of (\ref{Res.1}). Taking into account that addends
containing the non-zero powers of $\displaystyle
\frac{(\zeta-E_0)^{n}}{(\xi-E_0)^{n}(\xi-\zeta)}$ have an estimate
from above by ${\rm const}(|\xi|+1)^{-n}$ we get the result.

If $\kappa>\nu/4$, the kernel $G^{(\kappa)}(E)$ is Carleman in
virtue of Theorems~\ref{thmb}(2) and~\ref{Th1}. The arguments used
by the proof of \eqref{MT6} show that these kernels obey the required
properties (C1) and (C2).

(4) Taking into consideration the items (1) and (3) we
can prove the item (4) by the arguments used in the proof of the
statement \eqref{MT7}.

(5) This is an immediate consequence of the item (4).

(6) The continuity of the eigenfunctions of $H_{A,U}$
follows from the last statement in the item (3). Since
$\|e^{-tH_{A,U}}\|_{2,\infty}<\infty$ (see Theorem~\ref{Th1}), any
eigenfunction of $H_{A,U}$ is bounded.

(7) To get the derivative $\partial
G_{A,U}(x,y\,;\,\zeta)/\partial\zeta$ at a point
$\zeta_0\in\res{\, H_{A,U}}$, we use the expansion \eqref{Res.1}
with $E_0$ replaced by $\zeta_0$, and $\xi$ replaced by $H_{A,U}$.
Due to the item~(4), for sufficiently large $n$ the last term in
the right-hand side of~\eqref{Res.1} will have an integral kernel
which is uniformly bounded as $\zeta$ is in some small
neighborhood of $\zeta_0$. This proves the requested equality for
$k=1$. For $k>1$ one should consider the $k$-th powers in the both
sides of~\eqref{Res.1} and use the same arguments. \end{proof}

\section{Concluding remarks}\label{sec5}

It would be interesting to understand whether the estimates obtained admit
a generalization to the potentials from the Kato class on the manifold,
see~\eqref{AA8}. In this connection it would be also useful to know whether
the above definition of the Kato class is sufficient for these purposes
or one needs more restrictive conditions for the non-flat case. This question
is still open.

At the same time, we emphasize that the approach presented here works not only
to prove the continuity properties, but also allows
a more detailed analysis of the Green function. Let us mention one of possible
applications. In some problems connected with the renormalization
technique the asymptotic behavior near the diagonal $D$ is important.
Some corresponding estimates in the Euclidian space were proved
in~\cite{Sim}, in particular, in $L^2(\RR^3)$ the Green function
$G_V$ of $-\Delta+V$ with $V$ from the Kato class was shown to
satisfy the estimate
\[
\frac{C_1}{|x-y|}\le\big|G_V(x,y;\zeta)\big|\le\frac{C_2}{|x-y|}
\]
for small $|x-y|$ with some $C_1,C_2>0$.
Related propeties for  singular magnetic potentials are discussed
e.g.~in~\cite{Hab}. In~\cite{BGP} we represented
the Green function in lower dimensions ($\nu\le 3$)
in the form $G_{A,U}(x,y;\zeta)=F_{A,U}(x,y)+G_{A,U}^\text{ren}(x,y;\zeta)$,
where the second term on the right hand side  is continuous in the whole space $X\times X$,
and described the dependence of the singularity $F_{A,U}$
on the magnetic and electric potentials. It came out that this singularity may differ
from the standard one (fundamental solution for the Laplace operator) if the electric potential
becomes singular.

\section*{Acknowledgments}
We are grateful to S.~Albeverio, A.~Daletski, V.~Demidov,
M.~Demuth, P.~Exner, Yu.~Kordyukov, E.~Korotyaev, H.~Leschke, P.~M\"uller,
H.~Schulz-Baldes, and M.~Shubin for useful
discussions and valuable remarks.
The work was partially supported by the
Deutsche Forschungsgemeinschaft, INTAS, and TMR
HPRN-CT-1999-00118.

\end{document}